\documentclass[aps,prb,reprint,twocolumn,superscriptaddress]{revtex4-2}

\usepackage{graphicx,times,amsmath,xcolor,hyperref,stackengine}
\usepackage[caption=false]{subfig}
\hypersetup{colorlinks=true,final=true,linkcolor=blue,citecolor=blue,urlcolor=blue}
\graphicspath{{figs/}}

\newcommand{\ket}[1]{| {#1} \rangle}

\newcommand{\<}{\langle}

\renewcommand{\>}{\rangle}

\renewcommand{\bm}{\mathbf}

\begin{document}
	
	\title{Large circular photogalvanic effect in the noncentrosymmetric magnetic Weyl semimetal CeAlSi}

	\author{Abhirup Roy Karmakar}
	\email{abhirup.phy@iitkgp.ac.in}
	\altaffiliation{\\Present address: Dipartimento di Chimica,\\Università degli Studi di Milano, Via Golgi 19, 20133 Milano, Italy}
	\affiliation{Department of Physics, \href{https://www.iitkgp.ac.in}{Indian Institute of Technology Kharagpur}, Kharagpur 721302, India}
	
	\author{A. Taraphder}
	\email{arghya@phy.iitkgp.ac.in}
	\affiliation{Department of Physics, \href{https://www.iitkgp.ac.in}{Indian Institute of Technology Kharagpur}, Kharagpur 721302, India}
	
	\author{G. P. Das}
	\email{gour.das@tcgcrest.org}
	\affiliation{Research Institute for Sustainable Energy (RISE), \href{https://www.tcgcrest.org}{TCG-CREST}, Salt Lake, Kolkata 700091, India}

	\date{\today}

\begin{abstract}

	The recent discovery of the Weyl semimetal CeAlSi with simultaneous breaking of inversion and time-reversal symmetries has opened up new avenues for research into the interaction between light and topologically protected bands. In this work, we present a comprehensive examination of the shift current and injection current responsible for the circular photogalvanic effect in CeAlSi using first-principles calculations. Our investigation identifies a significant injection current of 4 mA/V$^2$ over a broad range in the near-infrared region of the electromagnetic spectrum, exceeding previously reported findings. In addition, we explored several externally controllable parameters to further enhance the photocurrent. A substantial boost in the injection current is observed when applying uniaxial strain along the $c$ axis of the crystal: a 5\% strain results in a remarkable 64\% increment. The exceptional photocurrent response in CeAlSi suggests that magnetic non-centrosymmetric Weyl semimetals may provide promising opportunities for novel photogalvanic applications. \\ \\
	DOI: \href{https://doi.org/10.1103/PhysRevB.111.195105}{10.1103/PhysRevB.111.195105}

\end{abstract}

	\maketitle

 \section{Introduction}\label{sec:introduction}

	Weyl semimetals have captivated the condensed matter physics community, drawing considerable attention due to their distinct electronic band structure and intriguing properties. These topological materials exhibit massless Weyl fermions as fundamental quasiparticles, in contrast to the conventional solid-state physics, and inspired cutting-edge research \cite{armitage_weyl_2018,bansil_colloquium_2016,yan_topological_2017}. The unique characteristics of Weyl semimetals (WSM) are closely linked to their symmetry-related features, making them a fascinating subject of study. In particular, Weyl semimetals break two critical symmetries in the crystal lattice: time-reversal symmetry (TRS) and inversion symmetry (IS) \cite{zyuzin_weyl_2012}. Time-reversal symmetry ensures the equivalence of particle dynamics when played backward in time, while inversion symmetry relates the system's behavior under spatial inversion. The breaking of these symmetries results in the formation of pairs of Weyl points in the electronic band structure. These Weyl points act as sources and sinks of quantized Berry curvature in momentum space \cite{xiao_berry_2010}. Additionally, the breaking of TRS in Weyl semimetals \cite{yang_quantum_2011,chang_topological_2018,roykarmakar_probing_2021} leads to the chiral anomaly — an astounding phenomenon resulting in an imbalance of chiral fermions in the presence of parallel electric and magnetic fields. This effect gives rise to exotic transport properties such as the negative longitudinal magnetoresistance \cite{huang_observation_2015,arnold_negative_2016,wang_gatetunable_2016} anomalous Hall effect \cite{burkov_anomalous_2014,steiner_anomalous_2017,shekhar_anomalous_2018}, anomalous thermal Hall effect \cite{roykarmakar_giant_2022,gorbar_anomalous_2017,ferreiros_anomalous_2017}, anomalous Nernst effect \cite{gorbar_anomalous_2017,ferreiros_anomalous_2017,watzman_dirac_2018} etc. The other class of WSMs with broken inversion symmetry \cite{lv_observation_2015,xu_experimental_2015,weng_weyl_2015,dey_dynamic_2020} produce another set of interesting electronic and optical phenomena like nonlinear anomalous Hall effect \cite{sodemann_quantum_2015,kumar_roomtemperature_2021,zeng_nonlinear_2021}, second harmonic generation \cite{lv_highharmonic_2021,lu_secondharmonic_2022,li_second_2018}, bulk photovoltaic effect \cite{ahn_lowfrequency_2020,osterhoudt_colossal_2019,tiwari_firstprinciples_2020}, circular photogalvanic effect \cite{dejuan_difference_2020,flicker_chiral_2018,sadhukhan_electronic_2021} etc.

	Among the remarkable manifestations in Weyl semimetals, the second-order optical responses have sparked a surge of interest in the field of topological materials due to their close connection with Berry phase. The bulk photovoltaic effect (BPVE) manifests as a nonlinear, optically-induced DC current generation within the material's bulk, arising from the breaking of inversion symmetry\cite{cook_design_2017,tan_shift_2016}. On the other hand, the circular photogalvanic effect (CPGE) is the generation of helicity-dependent current in non-centrosymmetric systems due to the irradiation of circularly polarized light \cite{belinicher_photogalvanic_1980,asnin_circular_1979,pikus_photogalvanic_1980}. 
	\begin{figure}[!hbt]
		\includegraphics[width=0.95\columnwidth]{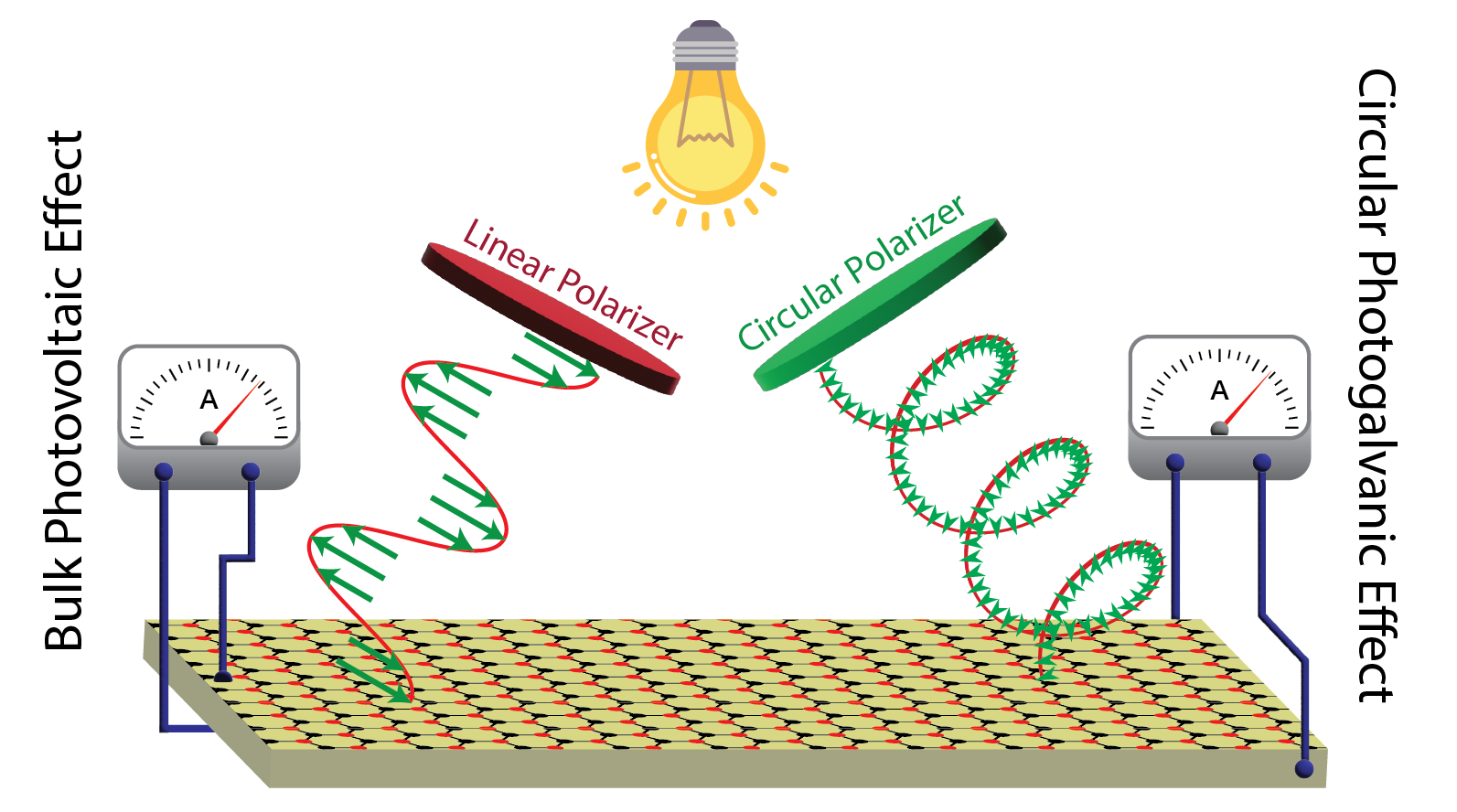}
		\caption{\textbf{Schematic diagram:} The left and right parts of the figure demonstrate the bulk photovoltaic effect and the circular photogalvanic effect respectively.}
		\label{fig:schematic}
	\end{figure}
	This gets more interesting when a WSM breaks mirror symmetry along with the inversion symmetry. Such materials exhibit quantized CPGE response which is directly proportional to the topological charge of the WSM and calculated as the trace of the CPGE tensor \cite{dejuan_quantized_2017,flicker_chiral_2018}. So far, BPVE and CPGE including the quantized responses have been extensively studied in Weyl semimetals both theoretically and experimentally \cite{tiwari_firstprinciples_2020,tiwari_firstprinciples_2020,osterhoudt_colossal_2019,zhang_photogalvanic_2018,wu_giant_2017,nag_distinct_2022,sessi_handednessdependent_2020,ibanez-azpiroz_initio_2018,dejuan_difference_2020,kumbhakar_reversible_2021,ni_giant_2021,ni_linear_2020,flicker_chiral_2018,dejuan_quantized_2017,le_initio_2020,nag_distinct_2022}. Efforts have been given to tune the materials by doping, changing chemical potential, applying electric fields and even increasing the hot-carrier scattering time in order to enhance the CPGE responses \cite{ni_giant_2021,ni_linear_2020,flicker_chiral_2018,le_initio_2020,wang_ferroicitydriven_2019}. Theoretical calculations have also been performed by considering three-band transitions instead of the generic two-band process to obtain larger injection current \cite{zhang_photogalvanic_2018,sadhukhan_electronic_2021}. Recently, theoretical investigations on nonlinear optical responses using model Hamiltonians have been carried out to understand the role of band topology, where they have predicted that injection current is enhanced in semimetals with broken TRS \cite{holder_consequences_2020,sadhukhan_role_2021}. However, the realistic Weyl semimetals wherein both inversion and time-reversal symmetries are concurrently broken remains limited and CPGE has not yet been studied in such materials.

	Notably, CeAlSi has been identified as a candidate of this category in the recent past, boasting noncollinear ferromagnetic order \cite{yang_noncollinear_2021}. While considerable research has been devoted to studying its anomalous transport properties and Fermi arcs \cite{yang_noncollinear_2021,alam_sign_2023,piva_topological_2023,sakhya_observation_2023,zhang_temperaturedependent_2023}, the material presents numerous uncharted areas of nonlinear optical activities. In this work on CeAlSi, we present a comprehensive first-principles investigation of the injection current, a key contributor to the circular photogalvanic effect, and the shift current that drives the bulk photovoltaic effect with a motivation to look for large photocurrent. Remarkably, our findings indicate a substantial magnitude of the injection current ($\sim$ 4 mA/V$^2$), surpassing previously reported literature results including those for non-magnetic Weyl semimetals \cite{ni_linear_2020,ni_giant_2021,zhang_photogalvanic_2018,dejuan_quantized_2017,le_initio_2020}. Intriguingly, it emerges as the dominant mechanism within the near-infrared region of the electromagnetic spectrum. In contrast, the contribution of the shift current remains comparatively limited with respect to the injection current. Moreover, our inquiry extends towards identifying several externally tunable parameters that positively influence the circular photocurrent. Interestingly, we observe a significant boost in the photocurrent through the application of uniaxial strain along the \textit{c}-axis of the crystal. This noteworthy revelation opens new avenues for the advancement of photovoltaic technologies and underscores the material's potential in light-driven applications.
	
	The following sections of this paper are arranged in the subsequent manner: Sec. \ref{sec:matherial_and_method} addresses the crystal structure, and the underlying symmetries of CeAlSi, along with an explanation of the computational methods utilized. In Sec. \ref{sec:electronic_structure}, a comprehensive analysis of its intrinsic magnetization, electronic structure, Weyl nodes and Berry curvature is conducted. Sec. \ref{sec:nonlinear_photocurrents} presents the theory and outcomes related to shift current and injection current. Subsequently, Sec. \ref{sec:external_influences} explores the impact of magnetic ordering, chemical potential and stress on the circular photocurrent and how it leads to enhancement. Finally, in Sec. \ref{sec:summary}, we offer a summary of our work and potential future research directions.

\section{Material and Method}\label{sec:matherial_and_method}

	\begin{figure}[!hbt]
		\includegraphics[width=6cm]{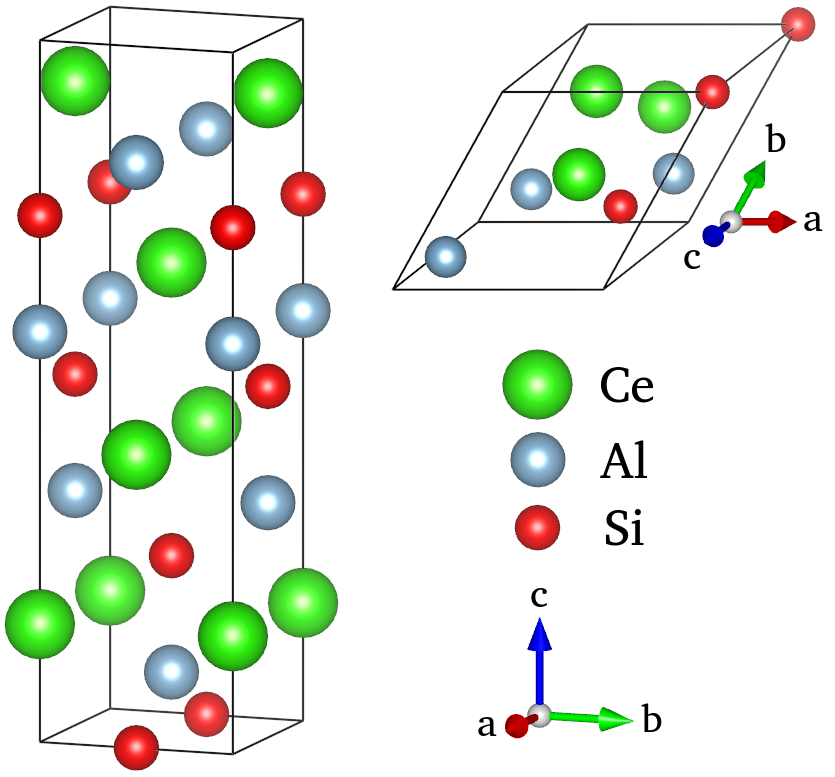}
		\caption{ \textbf{Crystal structure of CeAlSi.} The left part of the figure shows the tetragonal structure of the conventional unit cell (space group \textit{I4$_1$md}). The primitive cell of the system is shown on the right along with its lattice vectors. } 
		\label{fig:crystal_structure}
	\end{figure}

	 CeAlSi has a noncentrosymmetrically ordered structure with a space group of $I4_1md$ (No. 109) and is an isostructure of CeAlGe which was first proposed as a Ferromagnetic Weyl semimetal \cite{chang_magnetic_2018,xu_discovery_2017}. The tetragonal unit cell has the experimental lattice parameters $a=4.25$ \AA\ and $c=14.58$ \AA\ (Fig. \ref{fig:crystal_structure}). Although this crystal structure of point group $4mm$ possesses a four-fold rotation axis ($C_{4v}$) and two vertical mirror planes ($\sigma_v$), it lacks the horizontal mirror plane ($\sigma_h$) which is responsible for the breaking of inversion symmetry. On another note, site-mixing between the elements Al and Si could restore the horizontal mirror plane ($\sigma_h$) and thereby transforming the structure into a centrosymmetric one with space group $I4_1/amd$. However, people have confirmed the structure to be $I4_1md$ by measuring second-harmonic generation (SHG) which only occurs in non-centrosymmetric systems \cite{yang_noncollinear_2021}.

	We have performed first principles density functional theory (DFT) calculations using Vienna Ab initio Simulation Package (VASP) \cite{kresse_efficient_1996} in the projector augmented wave (PAW) approximation. Generalized gradient approximation (GGA) was considered for exchange-correlation functional in the Perdew-Burke-Ernzerhof (PBE) scheme \cite{perdew_generalized_1996}. A 15$\times$15$\times$15 Monk-horst grid was taken to fill the 3D Brillouin zone. To account for the $f$-electrons of Ce atoms, an on-site Coulomb interaction of strength $U_{eff}=$ 7 eV was added within the aproach given by Dudarev \textit{et al.} \cite{dudarev_electronenergyloss_1998}. We projected the Bloch wave functions obtained from this DFT+U calculation into the maximally localized Wannier functions (MLWFs) and obtained Hamiltonian matrix elements between the MLWFs with the help of Wannier90 package \cite{pizzi_wannier90_2020}. The \textit{d}-orbitals of Ce and \textit{s},\textit{p}-orbitals of Al and Si have been chosen as the projections for the Wannierization. In order for a thorough investigation of the band structure and calculation of the photocurrents, we derived a Hamiltonian using the Wannier tight-binding model with 52 bands from the matrix elements \cite{pizzi_wannier90_2020, gresch_automated_2018}. All other numerical calculations were performed in Python programming language and with the help of WannierBerri \cite{tsirkin_high_2021} package.

\section{Electronic structure}\label{sec:electronic_structure}

	\begin{figure*}[!hbt]
		\centering
		\includegraphics[width=0.9\textwidth]{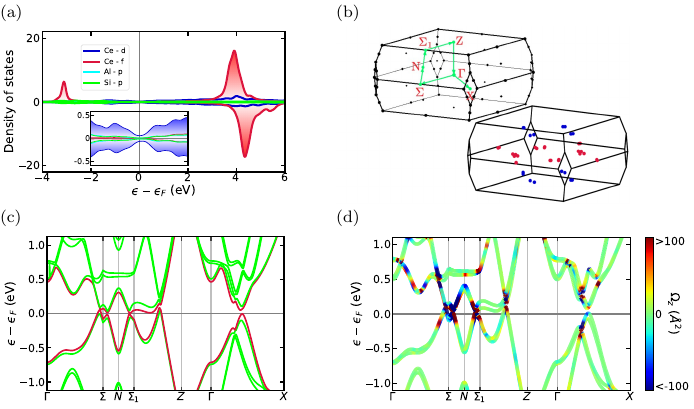}
		\caption{ \textbf{Electronic structure of CeAlSi.}
			\textbf{(a)} The orbital-projected density of states. The asymmetry in the large peaks of the red curve (Ce-\textit{f} orbital) indicates the magnetic nature of the system. However, the \textit{f}-orbital does not have much dominance near Fermi level as depicted in the inset figure. The $ + $ve and $ - $ve values correspond to spin-up and spin-down channels respectively.
			\textbf{(b)} Left: Brillouin zone of the unit cell where the green line indicates the high-symmetry path for band structure. Right: Locations of the Weyl nodes inside the Brillouin zone. Red dots are the nodes lying on the $k_z=0$ plane, whereas blue dots correspond to the rest.
			\textbf{(c)} The band structure with spin-orbit coupling is plotted along the high symmetry direction  $\Gamma-\Sigma-N-\Sigma_1-Z-\Gamma-X$. The valence and the conduction bands are colored in \textit{red}.
			\textbf{(d)} Berry curvature resolved band structure along the high-symmetry path. The color coding represents the values of the $z$-component of Berry curvature as shown in the color bar. }
		\label{fig:electronic_structure}
	\end{figure*}

	To start with, we plotted the orbital-projected density of states (PDOS) in Fig. \ref{fig:electronic_structure}a. Different colors corresponds to different orbitals as mentioned in the legend and positive and negative parts correspond to the spin-up and spin-down channels respectively. It is clearly visible that the \textit{f}-orbital of Ce peaks out the most and there is a significant asymmetry between the two spin channels. This dictates the magnetic nature of the system which predominantly comes only from the Ce-\textit{f} orbitals. The inset figure of PDOS is the zoomed-in version of the former one and gives us two important information : \textit{(i)} at energy levels close to the Fermi level ($\epsilon_F$), only the \textit{d}-orbital of Ce atoms dominate and the rest including Ce-\textit{f} have negligible contribution. \textit{(ii)} PDOS is symmetric near $\epsilon_F$ which implies that the magnetization of the system is very localized and away from $\epsilon_F$. Since the unit cell has element with large atomic number, it is customary to incorporate spin-orbit coupling (SOC). Upon performing the DFT calculation including SOC, we found that the system has a noncollinear Ferromagnetic order with net magnetization vector [1.1, 1.1, 0.3]. The individual Ce atoms in the unit cell, i.e. the major contributor to magnetization, has a magnetic moment of 0.89 $\mu_B$. Now, to gain more insight of the electronic structure, we plotted the band structure in Fig. \ref{fig:electronic_structure}c along the high-symmetry path $\Gamma-\Sigma-N-\Sigma_1-Z-\Gamma-X$ in the Brillouin zone as shown in Fig. \ref{fig:electronic_structure}b. The band-splittings are clearly visible in the plot which are caused by the effect of SOC. By observing the red lines representing the valence and conduction bands, one can see two band crossings around the high-symmetry points $\Sigma$ and one along the direction $N-\Sigma_1$. However, by scanning throughout the Brillouin zone in three-dimensional reciprocal space, we obtained 20 pairs of Weyl nodes in total. Their locations in the Brillouin zone are shown in Fig. \ref{fig:electronic_structure}b. Twelve pairs of them lie at the $k_y = 0$ plane while eight other pairs are located away from the plane. The breaking of inversion symmetry allows the Weyl nodes to be distributed over various energy levels ranging from 0 to 110 meV with respect to the Fermi level as shown in Fig. \ref{fig:bands-nodes_strain-z}d. It is important to note that the Weyl nodes are two-fold degenerate as only valence and conduction band participate in forming them. However, there exist pairs of nodes with equal energies due to the presence of mirror symmetry in the crystal (Fig. \ref{fig:bands-nodes_strain-z}d).

 \section{Nonlinear Photocurrents}\label{sec:nonlinear_photocurrents}

	Circular photogalvanic effect is the phenomenon where circularly polarized light incident on a material induces a time-dependent current density that manifests in a DC current response. Whereas, the bulk photovoltaic effect is a DC current response due to linearly polarized light. Although these two effects can be produced by various mechanisms, we focus here only on the second-order intrinsic contributions, namely the ``shift current" and ``injection current". The total nonlinear photocurrent density in terms of the two can be written as,
	\begin{equation}\label{eq:J_DC}
		 \<\bm J_{DC}\>^{(2)} = \<\bm J_{SC}\>^{(2)} + \<\bm J_{IC}\>^{(2)}
	\end{equation}
	where,
	\begin{align*}
		\<\bm J^a_{SC}\>^{(2)} = \sigma_2^{abc}(0; \omega, -\omega) \mathrm{E}^b(\omega) \mathrm{E}^c(-\omega) \\
		{\<\bm J^a_{IC}\>}^{(2)} = \tau \eta_2^{abc}(0; \omega, -\omega) \mathrm{E}^b(\omega) \mathrm{E}^c(-\omega)
	\end{align*}
	with $a,b,c$ being the cartesian indices. $\sigma_2^{abc}$ and $\eta_2^{abc}$ are respectively the shift-current and injection current susceptibility tensor. $\tau$ is the relaxation time os the system under consideration. The electric fields $\bm E(\omega)$ are real in case of linearly polarized light, whereas they are complex for circularly polarized light. For two-band transitions, the photocurrent tensors can be mathematically calculated using the following expressions \cite{sipe_secondorder_2000,zhang_switchable_2019}
	\begin{multline}\label{eq:sigma}
		\sigma_2^{abc} = -\frac{i \pi e^3}{2\hbar^2} \int \frac{d\bm k}{(2\pi)^3} \sum_{n,m} f_{nm}\ (r^{b}_{mn}r^{c}_{nm;a} + r^{c}_{mn}r^{b}_{nm;a}) \\
		\times \delta(\epsilon_{mn} - \hbar \omega)
	\end{multline}
	\begin{multline}\label{eq:eta}
		\eta_2^{abc} = -\frac{\pi e^3}{\hbar^2} \int \frac{d\bm k}{(2\pi)^3} \sum_{n,m} \Delta^a_{mn}\ f_{nm}\ (r^{b}_{mn}r^{c}_{nm}) \\ \times \delta(\epsilon_{mn} - \hbar \omega)
	\end{multline}
	where, $m,n$ are the band indices. $\epsilon_{mn} = \epsilon_m - \epsilon_n $, $ f_{mn} = f_m - f_n $ are the differences of band energies and Fermi-Dirac occupation number respectively. $ \Delta^a_{mn} = \partial_{k_a} \epsilon_{mn} $, $ r^{a}_{mn} = A^a_{mn} = i \<{m}\ket{\partial_{k_a} n} $ ($m \ne n$) is the inter-band matrix element (off-diagonal Berry connection), and $ r^{b}_{nm;a} = \partial_{k_a} r^{b}_{mn} - i(A^a_{nn}-A^a_{mm})r^{b}_{mn} $ is the generalized derivative of the matrix element. The response functions vanish in the presence of inversion symmetry as the functions are odd in $\bm k$. The complex term $(r^{b}_{mn}r^{c}_{nm})$ in Eq. (\ref{eq:eta}) can be broken down into a real and an imaginary part, where $Re(r^{b}_{mn} r^{c}_{nm}) = \frac{1}{2} \left\lbrace r^{b}_{mn}, r^{c}_{nm}\right\rbrace $ and $Im(r^{b}_{mn} r^{c}_{nm}) = \frac{1}{2} \left[r^{b}_{mn}, r^{c}_{nm}\right]$. When a system has time-reversal symmetry, only the imaginary part of $\eta_2$ survives. The real part vanishes, because TRS requires $ \eta_2 (\bm k) = - \eta^{*}_2 (- \bm k) $. Moreover, as $\eta^{(Im)}_2$ is anti-symmetric in the last two indices, which dictate the polarization of light, there is no injection current for linearly polarized light in time-reversal symmetric systems. These type of systems only exhibit circular photogalvanic effect. However, since our system has both IS and TRS broken, we also expect to see the linear photogalvanic effect (LPGE) which is the generation of injection current due to linearly polarized light. It arises from the real part of $\eta_2$ that is non-zero for our system \cite{drueke_nonlinear_2021, zhang_switchable_2019, holder_consequences_2020, kaplan_nonvanishing_2020, kaplan_unifying_2023}. The function $\delta(\epsilon_{mn} - \hbar \omega)$ represents the imaginary part of $(\epsilon_{mn} - \hbar\omega - i\hbar/\tau)^{-1}$ which appears in the general form of the photocurrent tensor \cite{zhang_switchable_2019}. The relaxation time ($\tau$) is also coming as a factor in total injection current as we are only considering transitions between two bands.
	\begin{figure}[!hbt]
		\centering
		\includegraphics[width=0.9\columnwidth]{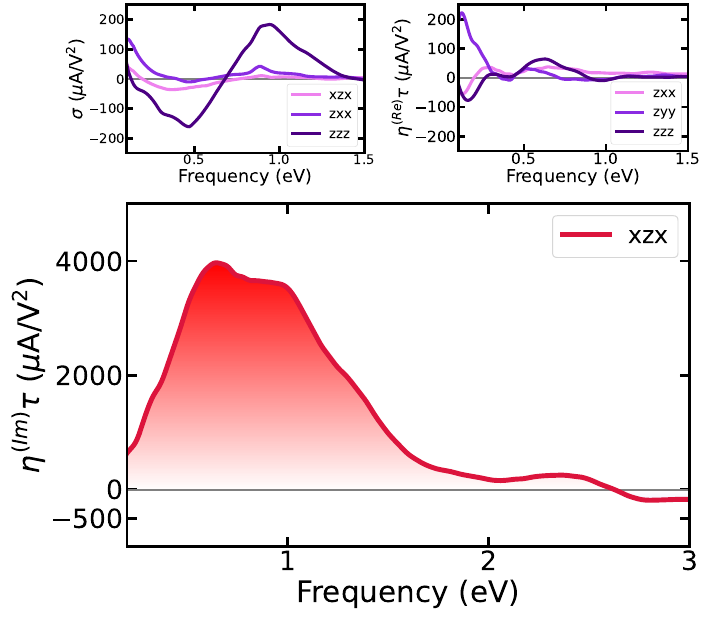}
		\caption{ Injection current and shift currents are plotted as a function of incident light frequency for $\frac{\hbar}{\tau} = 20$ meV. $\eta_{xzx}\tau$ has the maximum value of 3970 $\mu$A/V$^2$ and is dominant throughout a broad region in the infrared spectrum with the peak located at 0.64 eV. The intensity of the \textit{red} gradients is proportional to the corresponding values. On the other hand, LPGE and shift current values remain well below the above-mentioned value of CPGE current. }
		\label{fig:cpge-sc}
	\end{figure}

	The primary requirement for having non-zero values of the second order photocurrent tensors is the breaking of inversion symmetry which the material satisfies. Apart from that, any additional symmetry property of the crystal imposes restrictions on the form of the tensors. The diagonal components (e.g. $xyz=xx$) vanish due to the presence of mirror symmetries ($\sigma_v$) \cite{dejuan_quantized_2017}, which makes the system incapable of exhibiting quantized CPGE response. By explicit consideration of all the symmetries of the crystal class, one can determine the possible non-vanishing components out of the 27 components of $\sigma_2^{abc}$ and $\eta_2^{abc}$. \textit{Butcher} (1965) \cite{butcher_nonlinear_1965} performed those calculations for all the 32 crystallographic point groups, and the results are presented in Table 1.5.2 of Ref. \cite{boyd_nonlinear_2020}. The tetragonal crystal system of CeAlSi falls under the crystal class \textit{4mm} ($C_{4v}$). According to the table, the possible non-zero tensor elements for this material are $xzx = yzy$, $xxz = yyz$, $zxx$, $zyy$ and $zzz$ \cite{boyd_nonlinear_2020}. Additionally, the forms of the equations (\ref{eq:sigma}) and (\ref{eq:eta}) ensure that $\sigma_2^{xzx} = \sigma_2^{xxz}$ and $\eta_2^{xzx} = -\eta_2^{xxz}$. The first index represents the direction of photocurrent and the last two represent the polarization of light. Hence, both linearly- and circularly-polarized light are responsible for the generation of nonlinear photocurrents in CeAlSi. It is also obvious that swapping the helicity of the incident circularly-polarized light ($zx \leftrightarrow xz$) reverses the direction of the current, in accordance with the definition of CPGE.

	Since, the photocurrent tensors are obtained from transitions between multiple bands (both real and virtual), they are heavily dependent on the incident light frequency. Also, the relaxation time ($\tau$) plays a major role in the photocurrent calculation, especially for injection current. For a typical metallic system $\frac{\hbar}{\tau}$ should be of the order of $k_BT$. Here, we do our calculations in room temperature ($T = 300$ K) and accordingly choose $\frac{\hbar}{\tau} = 20$ meV ($\tau \sim  33 fs$). We integrate the quantities mentioned in the equations (\ref{eq:sigma}) and (\ref{eq:eta}) throughout the Brillouin zone considering transitions between all the 52 bands. The imaginary component ($\eta_2^{xzx}$) and real components of the injection current tensor along with the non-zero shift current components are plotted in Fig. \ref{fig:cpge-sc}. It is quite interesting to see that $\eta_2^{xzx}$ attains noticeably large values over a wide range in the electromagnetic spectrum with a maximum of 3970 $\mu$A/V$^2$, giving rise to a large circular photogalvanic effect. This current is induced along the \textit{a}-axis of the crystal when a right-circularly polarized light is irradiated along the \textit{b}-axis. Flipping the helicity of the light reverses the sign of the current. The peak is situated at 0.64 eV which correspond to wavelengths of 1940 $nm$, falling in the near-infrared (NIR) region ($2526-780 \ nm$). The plateau-like flat region between the peaks makes it more robust in terms of the frequency, which is only seen in quantized CPGE. This will allow to accumulate more current for a light source of wide range of frequencies. As per the full width at half maximum (FWHM), $\eta_{xzx}$ turns out to be dominant in the range $0.39 - 1.29$ eV ($3180 - 961 \ nm$). 
	\begin{figure}[!hbt]
		\centering
		\subfloat[0.65 eV]{
			\includegraphics[width=0.49\columnwidth]{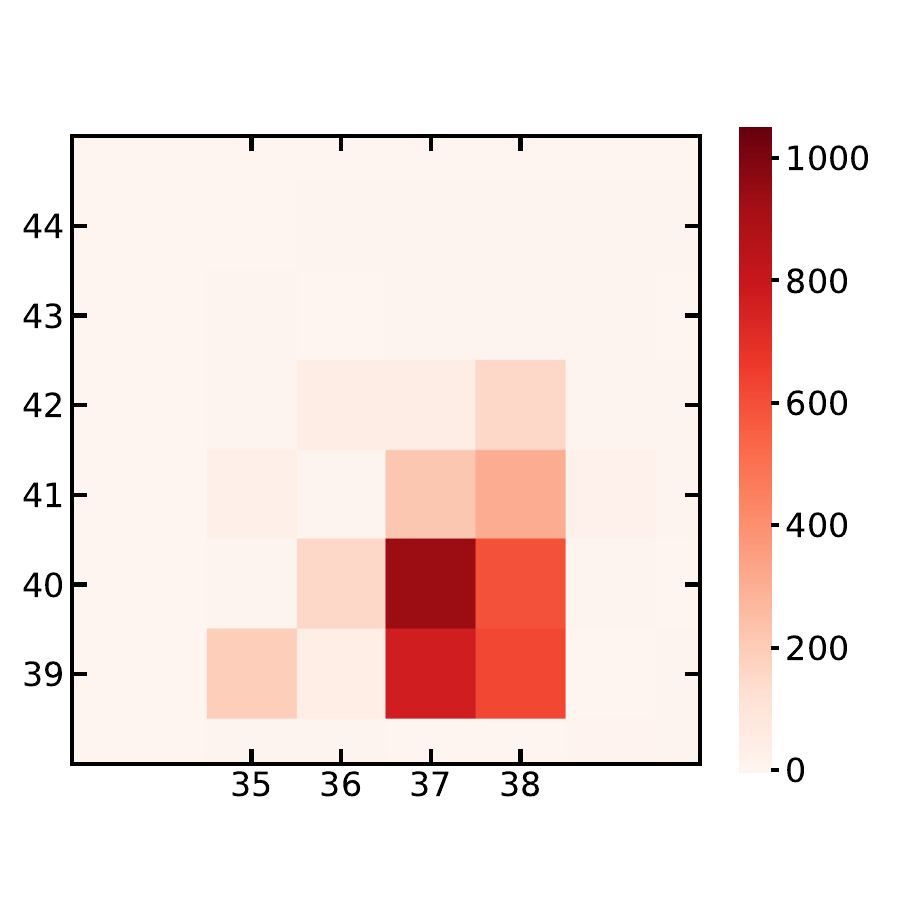}
			\label{fig:subfig1}
		}
		\subfloat[0.95 eV]{
			\includegraphics[width=0.49\columnwidth]{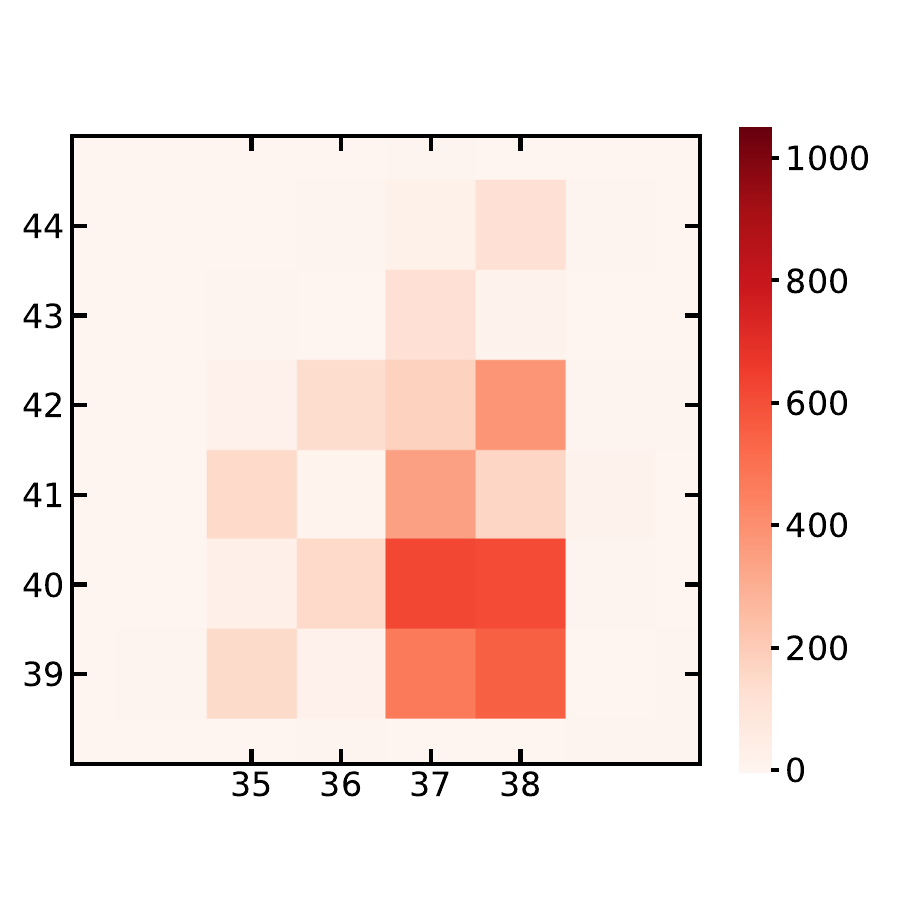}
			\label{fig:subfig2}
		}
		\caption{ Contribution of the bands leading to large $\eta_{xzx}$ at the plateau region.
			\textbf{(a)} Near the peak where the plateau starts (0.65 eV).
			\textbf{(b)} Near the right edge of the plateau (0.95) eV.
			In both cases, major contribution comes from the four bands around the Fermi level that have large Berry curvature.}
		\label{fig: IC_bands}
	\end{figure}
    To understand the source of the the plateau-like behavior of $\eta_{xzx}$, we plot the band contributions for that region in Fig. \ref{fig: IC_bands}. It turns out that, on both sides of the plateau, the major contribution to injection current comes from the four bands around the Fermi level. These are the bands that exhibit a large Berry curvature, as illustrated in Fig. \ref{fig:electronic_structure}d. However, as the frequency increases, these bands' contribution diminishes and disperses among a greater number of neighboring bands. In spite of that, the four bands that have large Berry curvature values dominate throughout the plateau region.
    Although the current decreases sharply beyond 1 eV, the contribution is still finite in the visible light area with 550 $\mu$A/V$^2$ at 1.63 eV (lower edge of the visible spectrum) and continues to decrease until 2.6 eV where where it reverses its polarity (positive to negative) and continues with small values. On the other hand, the LPGE contributions coming from the real part of $\eta_2$ (Fig. \ref{fig:cpge-sc}) remain quite low as compared to CPGE. The components of $\eta^{(Re)}_2$ goes only up to a value $\sim 100$ $\mu$A/V$^2$ in the finite frequency range as shown in the figure. The shift current which is induced by the separation of charges, also does not go as high as CPGE. The largest component $\sigma_{zzz}$ attains a maximum current of 185 $\mu$A/V$^2$ at 0.94 eV and has another negative peak of -155 $\mu$A/V$^2$ at 0.47 eV . It is produced along the \textit{c}-axis of the crystal when a linearly polarized light with polarization along the same direction falls on the material. The other two components have negligible response though. In the following part of the paper, we do further analysis of $\eta_{xzx}$ in presence of strain.

 \section{External Influences}\label{sec:external_influences}
	
\begin{figure}[!hbt]
		\centering
		\includegraphics[width=0.9\columnwidth]{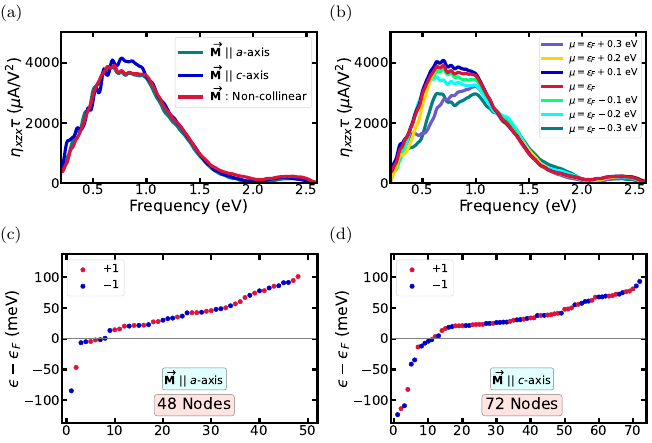}
		\caption{
			\textbf{(a)} Injection current plotted for different orientations of magnetic order. The photocurrent profile for magnetization along \textit{a}-axis (green) is almost similar to that of the noncollinear ground state (\textit{red}). However, the plateau region gets lifted when magnetization is along \textit{c}-axis (\textit{blue}), consequently shifting the initial peak position.
			\textbf{(b)} Multiple $\eta_{xzx}$ profiles plotted by varying chemical potential ($\mu$). It shows that the peak value increases only for a slight increase (up to 0.2 eV) in chemical potential from $\epsilon_F$. Increasing $\mu$ further or decreasing below $\epsilon_F$ lowers the value of $\eta_{xzx}$.
			\textbf{(c)}-\textbf{(d)} Distribution of Weyl nodes in energy landscape for magnetization along \textit{a}-axis and \textit{c}-axis respectively. The number of nodes and their degeneracies increased upon changing the magnetic ordering. Now band crossings occur below the Fermi level also, with energy as low as $-120$ meV in case of $\overrightarrow{\bm M}$ $||$ \textit{c}-axis.
		}
		\label{fig:cpge_mag-mu}
	\end{figure}

	Here, we look for the externally tunable parameters in order to maximize the current output. Motivated by the work of Alam \textit{et al.} \cite{alam_sign_2023}, we study CPGE for two different magnetic configurations. In its ground state, CeAlSi has noncollinear magnetic order with moments directed between \textit{a}-axis and \textit{b}-axis on \textit{ab}-plane. We calculate injection current for two cases of collinear magnetization: \textit{(i)} magnetic moments along \textit{a}-axis and \textit{(ii)} magnetic moments along \textit{c}-axis to examine if the magnetic symmetry plays any role to the current response. The ground state of CeAlSi has only two magnetic symmetry elements, namely the identity, two-fold screw rotation about $\left[1,1,0\right]$ direction. When the magnetization is constrained along $a$-axis, two more symmetry elements get unlocked. And when the magnetization is kept along the $c$-axis, there are 16 magnetic symmetry elements in total (including center translations) \cite{litvin_magnetic_2013,perez-mato_symmetrybased_2015,togo_spglib_2018}. We plot the corresponding results Fig. \ref{fig:cpge_mag-mu}a.
    
	\begin{figure}[!hbt]
		\centering
		\includegraphics[width=0.9\columnwidth]{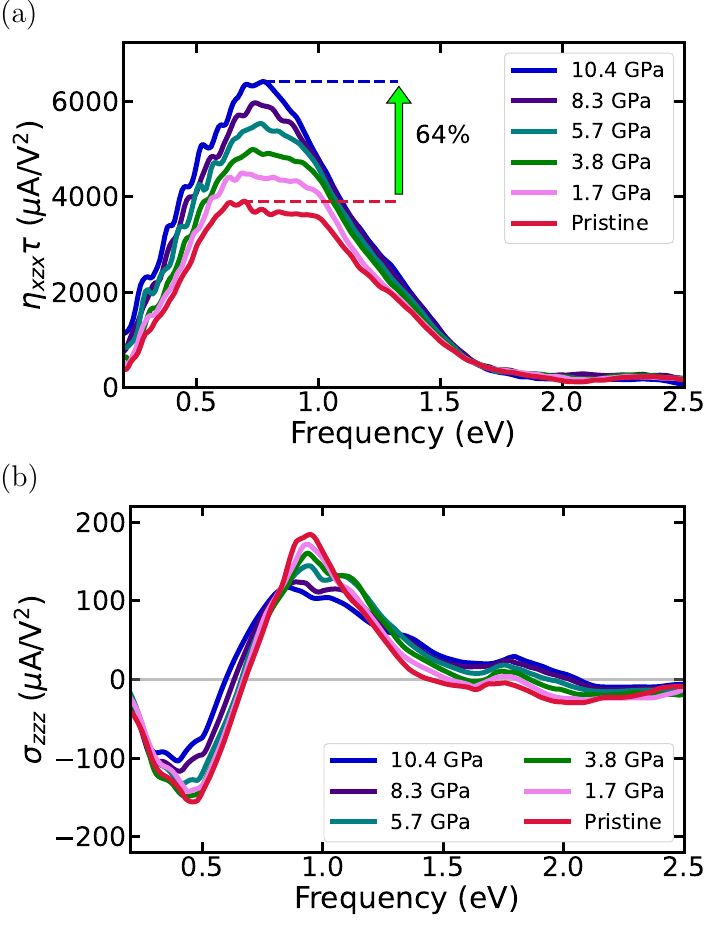}
		\caption{ The multicolored curves represent injection currents and shift currents for different amounts of compressive stress along \textit{c}-axis. The chemical potential was kept at the Femi level.
				\textbf{(a)} The values of injection current increase and the plateau region gradually peaks out at center with increasing stress. The maximum value reaches to 6510 $\mu$A/V$^2$ with an increment of 64\% at a stress of 10.4 GPa (\textit{blue}) from the ground state (\textit{red}).
				\textbf{(b)} Shift current values, however, show completely different behavior. The peaks get decreased when stress is applied. The current improves at higher frequencies though, where the values are much less than the peak values. }
		\label{fig:cpge_strain-z}
	\end{figure}
    
	\begin{figure*}[!hbt]
		\centering
		\includegraphics[width=0.9\textwidth]{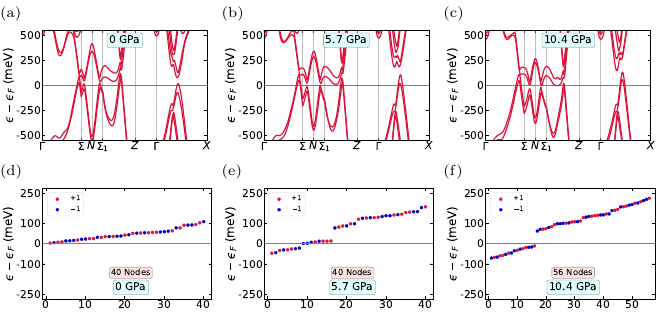}
		\caption{ \textbf{Strain-induced band dispersion and Weyl node energies.} \textbf{(a)}-\textbf{(c)} Band structures with SOC for different amounts of uniaxial compressive stress along \textit{c}-axis. The energy of the valence and conduction bands increases with stress along the paths $\Gamma - \Sigma$ and $\Gamma - X$ which are inside the Brillouin zone. However, the band energies decrease along the path $\Sigma-N-\Sigma_1-Z$ which lie on the outer surface of the Brillouin zone.
			\textbf{(d)}-\textbf{(f)} The Weyl nodes are plotted in energy scale and marked their chiralities with distinct colors (\textit{red} and \textit{blue}). The spread in energy increases with increasing stress and goes from 0 to 110 meV to a range between $-100$ and $250$ meV. Moreover, the number of nodes increases to 40 to 56 at a stress of 10.4 GPa. }
		\label{fig:bands-nodes_strain-z}
	\end{figure*}
	It turns out that when magnetization is along \textit{a}-axis, peak of the photocurrent profile remains almost similar to that of the ground state with noncollinear order. However, when the moments are forced to align towards the hard axis (\textit{c}-axis), the peak value increases by a small amount. The peak now shifts to the center of the plateau region and all the values in that region get enhanced. To be quantitative, the peak goes from 3970 $\mu$A/V$^2$ to 4150 $\mu$A/V$^2$. To understand this change in CPGE profile, we investigate the behavior of the Weyl nodes. We observe that the number of Weyl nodes increase when magnetic orientation is altered (Fig. \ref{fig:cpge_mag-mu}c-d). It becomes 48 when $\overrightarrow{\bm M}$ is along \textit{a}-axis and 72 for $\overrightarrow{\bm M}$ parallel to \textit{c}-axis as compared to 40 at ground state. Some of them appear below the Fermi level with energy as low as $-120$ meV. Moreover, the horizontal distributions of the nodes dictates that the degeneracy in energy also increases due to this transition. One should also note that, we didn’t observe any significant changes in other components of $\eta$. While there are slight variations due to changes in spin-orbit interactions, no new dominant component emerges, except for $\eta_{xzx}$. It turns out that the magnetic symmetry does not influence the overall behavior of the tensor. However, due to the changes in microscopic interactions with magnetic moments, we could observe some improvement in the dominant component. To proceed with another approach, we change the chemical potential ($\mu$) of the system and study the injection current profile, for multiple values of $\mu$ between $-0.3$ eV and $+0.3$ eV relative to the Fermi level (Fig. \ref{fig:cpge_mag-mu}b). We observe that, when we increase the chemical potential above $\epsilon_F$, $\eta_{xzx}$ starts to increase slightly at first. However, increasing $\mu$ beyond 0.2 eV (critical point), makes the current fall in the plateau region. Nevertheless, $\eta_{xzx}$ always shows a declining nature when decreasing $\mu$. Therefore, the two approaches mentioned above barely serve our purpose of enhancing the circular photogalvanic effect.

	To account for yet another external influence, we choose strain as the parameter. We apply compressive strain on the unit cell in two difference ways: \textit{(i)} along \textit{a}-axis and \textit{(ii)} along \textit{c}-axis. In the case of \textit{(i)}, we observe a steep decline in the magnitude of $\eta_{xzx}$. Whereas, applying compressive stress along the \textit{c}-axis drastically improves the circular photogalvanic effect. We start with the relaxed system in its ground state and gradually decrease the length of \textit{c}-axis in steps of 1\%. The photocurrent profiles for all those strain configurations are depicted in  Fig. \ref{fig:cpge_strain-z}a with the legends implying the stresses corresponding to each step. The maximum value of $\eta_{xzx}$ goes up by 64\% from the pristine value at 5\% strain corresponding to a stress of 10.4 GPa; which implies a huge current response of 6510 $\mu$A/V$^2$. Not only that, the plateau region of the \textit{red} curve gets sharpened gradually with the application of stress and the peak gets centered in the FWHM region. This leads to a shift of the peak frequency from 0.64 eV to 0.77 eV for a stress of 10.4 GPa. It is important to note that the atomic positions were relaxed after each application of stress, and the chemical potential was kept at the Fermi level ($\epsilon_F$). We have also tried changing the chemical potential along with the stress, but could not observe any significant improvement. Furthermore, we have studied the shift currents under strain (Fig. \ref{fig:cpge_strain-z}b). However, unlike the injection current, $\sigma_{zzz}$ peaks decreased when the stresses are applied. The values away from the peaks are somewhat increased as compared to the pristine, but still remain very small relative to peak values.
	
    To understand the effect of strain on shift-current and injection current, we carefully analyze Eq. (\ref{eq:sigma}) and (\ref{eq:eta}). The term $r^c_{nm;a}$ in the expression of $\sigma_2^{abc}$, which is defined as the generalized derivative of the Berry connection, is directly proportional to the shift vector. Shift vector is defined as
    $
    \mathcal{R}^b_{mn;a} = \partial_{k_a} \text{Arg}(A^b_{mn}) + (A^a_{mm} - A^a_{nn})
    $. This vector quantifies the real-space displacement of the electrons during optical transitions \cite{tan_effect_2019,kaner_enhanced_2020}. Localized wavefunctions cause a decrease in the shift vector, while delocalized wavefunctions do exactly the opposite. Consequently, the bunching of the wave packets under compressive strain leads to a decrease in shift current. The injection current, however, is primarily influenced by the diagonal components of the band velocities, or more precisely, the difference between the band velocities ($\Delta^a_{mn}$). The observed enhancement of the $xzx$-component of the injection current is due to the Poisson effect. Specifically, the uniaxial compressive strain along the $z$-direction leads to a tensile strain in the $x$-direction. This in turn results in an overall increase in $\Delta^x_{mn}$ for the dominant band transitions (Fig. \ref{fig: IC_bands}). Therefore, we realize the increase in the injection current under compressive strain.Another interesting point to look at is the distribution of Weyl nodes Figs. \ref{fig:bands-nodes_strain-z}(d-f). The system in its ground state holds 20 pairs of nodes with opposite chirality having energy up to $110$ meV. At 5.7 GPa stress, the nodes scatter more in energy scale and remain between $-50$ to $180$ meV. Interestingly, increasing stress beyond this point creates more Weyl points and at a stress of 10.4 GPa, there are 56 nodes in total scattered between $-100$ and $250$ meV. In addition, the degeneracies of the nodes also increase with pressure, which is indicated by the horizontal distribution of the nodes in energy landscape.

 \section{Summary}\label{sec:summary}

	We have presented our findings on nonlinear photocurrents in the non-centrosymmetric WSM CeAlSi induced by both linearly and circularly polarized light, using a multi-band approach, derived from the first-principles calculations. Our investigation commenced by analyzing the electronic structure and Berry curvature of CeAlSi. We observed that the material possesses a noncollinear ferromagnetic nature due to the \textit{f}-electron of Ce atom, thereby breaking the time-reversal symmetry. Moreover, the bands exhibit relatively high magnitudes of Berry curvature. On integration of relevant quantities as described in equations (\ref{eq:sigma}) and (\ref{eq:eta}) across 52 bands throughout the Brillouin zone, we were able to determine the shift current responsible for the bulk photovoltaic effect, as well as the injection current driving the circular photogalvanic effect. The results implied that this magnetic WSM draws significantly large injection current on irradiation of light compared as compared to time-reversal symmetric WSMs that were investigated before. Notably, the dominant portion of the injection photocurrent is located within the near-infrared (NIR) region of the electromagnetic spectrum. Additionally, we  explored various external parameters to modulate the intrinsic photocurrent response. Our findings indicate that a uniaxial compressive strain along the \textit{c}-axis gives a significant enhancement of the photocurrent. For instance, a stress of 10.4 GPa resulted in a remarkable 64\% increase in the injection current compared to the ground state value. Upon analyzing the electronic structure under strain, we discovered that the increase in band velocities is the reason behind this amplification of the injection current. In conclusion, our work suggests that the Weyl semimetal CeAlSi holds great promise for photonic applications as well as a renewable energy source, contributing positively to environmental sustainability.

 \begin{acknowledgments}
	A.R.K. acknowledges MHRD, Government of India for a research fellowship. We also acknowledge National Supercomputing Mission (NSM) for providing computing resources of ``PARAM Shakti'' at IIT Kharagpur, which is implemented by C-DAC and supported by the Ministry of Electronics and Information Technology (MeitY) and Department of Science and Technology (DST), Government of India.
 \end{acknowledgments}


 \bibliography{citations}

\begin{thebibliography}{77}%
\makeatletter
\providecommand \@ifxundefined [1]{%
 \@ifx{#1\undefined}
}%
\providecommand \@ifnum [1]{%
 \ifnum #1\expandafter \@firstoftwo
 \else \expandafter \@secondoftwo
 \fi
}%
\providecommand \@ifx [1]{%
 \ifx #1\expandafter \@firstoftwo
 \else \expandafter \@secondoftwo
 \fi
}%
\providecommand \natexlab [1]{#1}%
\providecommand \enquote  [1]{``#1''}%
\providecommand \bibnamefont  [1]{#1}%
\providecommand \bibfnamefont [1]{#1}%
\providecommand \citenamefont [1]{#1}%
\providecommand \href@noop [0]{\@secondoftwo}%
\providecommand \href [0]{\begingroup \@sanitize@url \@href}%
\providecommand \@href[1]{\@@startlink{#1}\@@href}%
\providecommand \@@href[1]{\endgroup#1\@@endlink}%
\providecommand \@sanitize@url [0]{\catcode `\\12\catcode `\$12\catcode
  `\&12\catcode `\#12\catcode `\^12\catcode `\_12\catcode `\%12\relax}%
\providecommand \@@startlink[1]{}%
\providecommand \@@endlink[0]{}%
\providecommand \url  [0]{\begingroup\@sanitize@url \@url }%
\providecommand \@url [1]{\endgroup\@href {#1}{\urlprefix }}%
\providecommand \urlprefix  [0]{URL }%
\providecommand \Eprint [0]{\href }%
\providecommand \doibase [0]{https://doi.org/}%
\providecommand \selectlanguage [0]{\@gobble}%
\providecommand \bibinfo  [0]{\@secondoftwo}%
\providecommand \bibfield  [0]{\@secondoftwo}%
\providecommand \translation [1]{[#1]}%
\providecommand \BibitemOpen [0]{}%
\providecommand \bibitemStop [0]{}%
\providecommand \bibitemNoStop [0]{.\EOS\space}%
\providecommand \EOS [0]{\spacefactor3000\relax}%
\providecommand \BibitemShut  [1]{\csname bibitem#1\endcsname}%
\let\auto@bib@innerbib\@empty
\bibitem [{\citenamefont {Armitage}\ \emph {et~al.}(2018)\citenamefont
  {Armitage}, \citenamefont {Mele},\ and\ \citenamefont
  {Vishwanath}}]{armitage_weyl_2018}%
  \BibitemOpen
  \bibfield  {author} {\bibinfo {author} {\bibfnamefont {N.~P.}\ \bibnamefont
  {Armitage}}, \bibinfo {author} {\bibfnamefont {E.~J.}\ \bibnamefont {Mele}},\
  and\ \bibinfo {author} {\bibfnamefont {A.}~\bibnamefont {Vishwanath}},\
  }\bibfield  {title} {\bibinfo {title} {Weyl and {{Dirac}} semimetals in
  three-dimensional solids},\ }\href
  {https://doi.org/10.1103/RevModPhys.90.015001} {\bibfield  {journal}
  {\bibinfo  {journal} {Rev. Mod. Phys.}\ }\textbf {\bibinfo {volume} {90}},\
  \bibinfo {pages} {015001} (\bibinfo {year} {2018})}\BibitemShut {NoStop}%
\bibitem [{\citenamefont {Bansil}\ \emph {et~al.}(2016)\citenamefont {Bansil},
  \citenamefont {Lin},\ and\ \citenamefont {Das}}]{bansil_colloquium_2016}%
  \BibitemOpen
  \bibfield  {author} {\bibinfo {author} {\bibfnamefont {A.}~\bibnamefont
  {Bansil}}, \bibinfo {author} {\bibfnamefont {H.}~\bibnamefont {Lin}},\ and\
  \bibinfo {author} {\bibfnamefont {T.}~\bibnamefont {Das}},\ }\bibfield
  {title} {\bibinfo {title} {{\emph{Colloquium}} : {{Topological}} band
  theory},\ }\href {https://doi.org/10.1103/RevModPhys.88.021004} {\bibfield
  {journal} {\bibinfo  {journal} {Rev. Mod. Phys.}\ }\textbf {\bibinfo {volume}
  {88}},\ \bibinfo {pages} {021004} (\bibinfo {year} {2016})}\BibitemShut
  {NoStop}%
\bibitem [{\citenamefont {Yan}\ and\ \citenamefont
  {Felser}(2017)}]{yan_topological_2017}%
  \BibitemOpen
  \bibfield  {author} {\bibinfo {author} {\bibfnamefont {B.}~\bibnamefont
  {Yan}}\ and\ \bibinfo {author} {\bibfnamefont {C.}~\bibnamefont {Felser}},\
  }\bibfield  {title} {\bibinfo {title} {Topological {{Materials}}: {{Weyl
  Semimetals}}},\ }\href
  {https://doi.org/10.1146/annurev-conmatphys-031016-025458} {\bibfield
  {journal} {\bibinfo  {journal} {Annu. Rev. Condens. Matter Phys.}\ }\textbf
  {\bibinfo {volume} {8}},\ \bibinfo {pages} {337} (\bibinfo {year}
  {2017})}\BibitemShut {NoStop}%
\bibitem [{\citenamefont {Zyuzin}\ \emph {et~al.}(2012)\citenamefont {Zyuzin},
  \citenamefont {Wu},\ and\ \citenamefont {Burkov}}]{zyuzin_weyl_2012}%
  \BibitemOpen
  \bibfield  {author} {\bibinfo {author} {\bibfnamefont {A.~A.}\ \bibnamefont
  {Zyuzin}}, \bibinfo {author} {\bibfnamefont {S.}~\bibnamefont {Wu}},\ and\
  \bibinfo {author} {\bibfnamefont {A.~A.}\ \bibnamefont {Burkov}},\ }\bibfield
   {title} {\bibinfo {title} {Weyl semimetal with broken time reversal and
  inversion symmetries},\ }\href {https://doi.org/10.1103/PhysRevB.85.165110}
  {\bibfield  {journal} {\bibinfo  {journal} {Phys. Rev. B}\ }\textbf {\bibinfo
  {volume} {85}},\ \bibinfo {pages} {165110} (\bibinfo {year}
  {2012})}\BibitemShut {NoStop}%
\bibitem [{\citenamefont {Xiao}\ \emph {et~al.}(2010)\citenamefont {Xiao},
  \citenamefont {Chang},\ and\ \citenamefont {Niu}}]{xiao_berry_2010}%
  \BibitemOpen
  \bibfield  {author} {\bibinfo {author} {\bibfnamefont {D.}~\bibnamefont
  {Xiao}}, \bibinfo {author} {\bibfnamefont {M.-C.}\ \bibnamefont {Chang}},\
  and\ \bibinfo {author} {\bibfnamefont {Q.}~\bibnamefont {Niu}},\ }\bibfield
  {title} {\bibinfo {title} {Berry phase effects on electronic properties},\
  }\href {https://doi.org/10.1103/RevModPhys.82.1959} {\bibfield  {journal}
  {\bibinfo  {journal} {Rev. Mod. Phys.}\ }\textbf {\bibinfo {volume} {82}},\
  \bibinfo {pages} {1959} (\bibinfo {year} {2010})}\BibitemShut {NoStop}%
\bibitem [{\citenamefont {Yang}\ \emph {et~al.}(2011)\citenamefont {Yang},
  \citenamefont {Lu},\ and\ \citenamefont {Ran}}]{yang_quantum_2011}%
  \BibitemOpen
  \bibfield  {author} {\bibinfo {author} {\bibfnamefont {K.-Y.}\ \bibnamefont
  {Yang}}, \bibinfo {author} {\bibfnamefont {Y.-M.}\ \bibnamefont {Lu}},\ and\
  \bibinfo {author} {\bibfnamefont {Y.}~\bibnamefont {Ran}},\ }\bibfield
  {title} {\bibinfo {title} {Quantum {{Hall}} effects in a {{Weyl}} semimetal:
  {{Possible}} application in pyrochlore iridates},\ }\href
  {https://doi.org/10.1103/PhysRevB.84.075129} {\bibfield  {journal} {\bibinfo
  {journal} {Phys. Rev. B}\ }\textbf {\bibinfo {volume} {84}},\ \bibinfo
  {pages} {075129} (\bibinfo {year} {2011})}\BibitemShut {NoStop}%
\bibitem [{\citenamefont {Chang}\ \emph
  {et~al.}(2018{\natexlab{a}})\citenamefont {Chang}, \citenamefont {Wieder},
  \citenamefont {Schindler}, \citenamefont {Sanchez}, \citenamefont
  {Belopolski}, \citenamefont {Huang}, \citenamefont {Singh}, \citenamefont
  {Wu}, \citenamefont {Chang}, \citenamefont {Neupert}, \citenamefont {Xu},
  \citenamefont {Lin},\ and\ \citenamefont {Hasan}}]{chang_topological_2018}%
  \BibitemOpen
  \bibfield  {author} {\bibinfo {author} {\bibfnamefont {G.}~\bibnamefont
  {Chang}}, \bibinfo {author} {\bibfnamefont {B.~J.}\ \bibnamefont {Wieder}},
  \bibinfo {author} {\bibfnamefont {F.}~\bibnamefont {Schindler}}, \bibinfo
  {author} {\bibfnamefont {D.~S.}\ \bibnamefont {Sanchez}}, \bibinfo {author}
  {\bibfnamefont {I.}~\bibnamefont {Belopolski}}, \bibinfo {author}
  {\bibfnamefont {S.-M.}\ \bibnamefont {Huang}}, \bibinfo {author}
  {\bibfnamefont {B.}~\bibnamefont {Singh}}, \bibinfo {author} {\bibfnamefont
  {D.}~\bibnamefont {Wu}}, \bibinfo {author} {\bibfnamefont {T.-R.}\
  \bibnamefont {Chang}}, \bibinfo {author} {\bibfnamefont {T.}~\bibnamefont
  {Neupert}}, \bibinfo {author} {\bibfnamefont {S.-Y.}\ \bibnamefont {Xu}},
  \bibinfo {author} {\bibfnamefont {H.}~\bibnamefont {Lin}},\ and\ \bibinfo
  {author} {\bibfnamefont {M.~Z.}\ \bibnamefont {Hasan}},\ }\bibfield  {title}
  {\bibinfo {title} {Topological quantum properties of chiral crystals},\
  }\href {https://doi.org/10.1038/s41563-018-0169-3} {\bibfield  {journal}
  {\bibinfo  {journal} {Nature Mater}\ }\textbf {\bibinfo {volume} {17}},\
  \bibinfo {pages} {978} (\bibinfo {year} {2018}{\natexlab{a}})}\BibitemShut
  {NoStop}%
\bibitem [{\citenamefont {Roy~Karmakar}\ \emph {et~al.}(2021)\citenamefont
  {Roy~Karmakar}, \citenamefont {Nandy}, \citenamefont {Das},\ and\
  \citenamefont {Saha}}]{roykarmakar_probing_2021}%
  \BibitemOpen
  \bibfield  {author} {\bibinfo {author} {\bibfnamefont {A.}~\bibnamefont
  {Roy~Karmakar}}, \bibinfo {author} {\bibfnamefont {S.}~\bibnamefont {Nandy}},
  \bibinfo {author} {\bibfnamefont {G.~P.}\ \bibnamefont {Das}},\ and\ \bibinfo
  {author} {\bibfnamefont {K.}~\bibnamefont {Saha}},\ }\bibfield  {title}
  {\bibinfo {title} {Probing mirror anomaly and classes of {{Dirac}} semimetals
  with circular dichroism},\ }\href
  {https://doi.org/10.1103/PhysRevResearch.3.013230} {\bibfield  {journal}
  {\bibinfo  {journal} {Phys. Rev. Research}\ }\textbf {\bibinfo {volume}
  {3}},\ \bibinfo {pages} {013230} (\bibinfo {year} {2021})}\BibitemShut
  {NoStop}%
\bibitem [{\citenamefont {Huang}\ \emph {et~al.}(2015)\citenamefont {Huang},
  \citenamefont {Zhao}, \citenamefont {Long}, \citenamefont {Wang},
  \citenamefont {Chen}, \citenamefont {Yang}, \citenamefont {Liang},
  \citenamefont {Xue}, \citenamefont {Weng}, \citenamefont {Fang},
  \citenamefont {Dai},\ and\ \citenamefont {Chen}}]{huang_observation_2015}%
  \BibitemOpen
  \bibfield  {author} {\bibinfo {author} {\bibfnamefont {X.}~\bibnamefont
  {Huang}}, \bibinfo {author} {\bibfnamefont {L.}~\bibnamefont {Zhao}},
  \bibinfo {author} {\bibfnamefont {Y.}~\bibnamefont {Long}}, \bibinfo {author}
  {\bibfnamefont {P.}~\bibnamefont {Wang}}, \bibinfo {author} {\bibfnamefont
  {D.}~\bibnamefont {Chen}}, \bibinfo {author} {\bibfnamefont {Z.}~\bibnamefont
  {Yang}}, \bibinfo {author} {\bibfnamefont {H.}~\bibnamefont {Liang}},
  \bibinfo {author} {\bibfnamefont {M.}~\bibnamefont {Xue}}, \bibinfo {author}
  {\bibfnamefont {H.}~\bibnamefont {Weng}}, \bibinfo {author} {\bibfnamefont
  {Z.}~\bibnamefont {Fang}}, \bibinfo {author} {\bibfnamefont {X.}~\bibnamefont
  {Dai}},\ and\ \bibinfo {author} {\bibfnamefont {G.}~\bibnamefont {Chen}},\
  }\bibfield  {title} {\bibinfo {title} {Observation of the
  {{Chiral-Anomaly-Induced Negative Magnetoresistance}} in {{3D Weyl Semimetal
  TaAs}}},\ }\href {https://doi.org/10.1103/PhysRevX.5.031023} {\bibfield
  {journal} {\bibinfo  {journal} {Phys. Rev. X}\ }\textbf {\bibinfo {volume}
  {5}},\ \bibinfo {pages} {031023} (\bibinfo {year} {2015})}\BibitemShut
  {NoStop}%
\bibitem [{\citenamefont {Arnold}\ \emph {et~al.}(2016)\citenamefont {Arnold},
  \citenamefont {Shekhar}, \citenamefont {Wu}, \citenamefont {Sun},
  \citenamefont {Dos~Reis}, \citenamefont {Kumar}, \citenamefont {Naumann},
  \citenamefont {Ajeesh}, \citenamefont {Schmidt}, \citenamefont {Grushin},
  \citenamefont {Bardarson}, \citenamefont {Baenitz}, \citenamefont {Sokolov},
  \citenamefont {Borrmann}, \citenamefont {Nicklas}, \citenamefont {Felser},
  \citenamefont {Hassinger},\ and\ \citenamefont {Yan}}]{arnold_negative_2016}%
  \BibitemOpen
  \bibfield  {author} {\bibinfo {author} {\bibfnamefont {F.}~\bibnamefont
  {Arnold}}, \bibinfo {author} {\bibfnamefont {C.}~\bibnamefont {Shekhar}},
  \bibinfo {author} {\bibfnamefont {S.-C.}\ \bibnamefont {Wu}}, \bibinfo
  {author} {\bibfnamefont {Y.}~\bibnamefont {Sun}}, \bibinfo {author}
  {\bibfnamefont {R.~D.}\ \bibnamefont {Dos~Reis}}, \bibinfo {author}
  {\bibfnamefont {N.}~\bibnamefont {Kumar}}, \bibinfo {author} {\bibfnamefont
  {M.}~\bibnamefont {Naumann}}, \bibinfo {author} {\bibfnamefont {M.~O.}\
  \bibnamefont {Ajeesh}}, \bibinfo {author} {\bibfnamefont {M.}~\bibnamefont
  {Schmidt}}, \bibinfo {author} {\bibfnamefont {A.~G.}\ \bibnamefont
  {Grushin}}, \bibinfo {author} {\bibfnamefont {J.~H.}\ \bibnamefont
  {Bardarson}}, \bibinfo {author} {\bibfnamefont {M.}~\bibnamefont {Baenitz}},
  \bibinfo {author} {\bibfnamefont {D.}~\bibnamefont {Sokolov}}, \bibinfo
  {author} {\bibfnamefont {H.}~\bibnamefont {Borrmann}}, \bibinfo {author}
  {\bibfnamefont {M.}~\bibnamefont {Nicklas}}, \bibinfo {author} {\bibfnamefont
  {C.}~\bibnamefont {Felser}}, \bibinfo {author} {\bibfnamefont
  {E.}~\bibnamefont {Hassinger}},\ and\ \bibinfo {author} {\bibfnamefont
  {B.}~\bibnamefont {Yan}},\ }\bibfield  {title} {\bibinfo {title} {Negative
  magnetoresistance without well-defined chirality in the {{Weyl}} semimetal
  {{TaP}}},\ }\href {https://doi.org/10.1038/ncomms11615} {\bibfield  {journal}
  {\bibinfo  {journal} {Nat Commun}\ }\textbf {\bibinfo {volume} {7}},\
  \bibinfo {pages} {11615} (\bibinfo {year} {2016})}\BibitemShut {NoStop}%
\bibitem [{\citenamefont {Wang}\ \emph {et~al.}(2016)\citenamefont {Wang},
  \citenamefont {Liu}, \citenamefont {Liu}, \citenamefont {Pan}, \citenamefont
  {Zhang}, \citenamefont {Zeng}, \citenamefont {Fu}, \citenamefont {Wang},
  \citenamefont {Xu}, \citenamefont {Huang}, \citenamefont {Wang},
  \citenamefont {Lu}, \citenamefont {Xing}, \citenamefont {Wang}, \citenamefont
  {Wan},\ and\ \citenamefont {Miao}}]{wang_gatetunable_2016}%
  \BibitemOpen
  \bibfield  {author} {\bibinfo {author} {\bibfnamefont {Y.}~\bibnamefont
  {Wang}}, \bibinfo {author} {\bibfnamefont {E.}~\bibnamefont {Liu}}, \bibinfo
  {author} {\bibfnamefont {H.}~\bibnamefont {Liu}}, \bibinfo {author}
  {\bibfnamefont {Y.}~\bibnamefont {Pan}}, \bibinfo {author} {\bibfnamefont
  {L.}~\bibnamefont {Zhang}}, \bibinfo {author} {\bibfnamefont
  {J.}~\bibnamefont {Zeng}}, \bibinfo {author} {\bibfnamefont {Y.}~\bibnamefont
  {Fu}}, \bibinfo {author} {\bibfnamefont {M.}~\bibnamefont {Wang}}, \bibinfo
  {author} {\bibfnamefont {K.}~\bibnamefont {Xu}}, \bibinfo {author}
  {\bibfnamefont {Z.}~\bibnamefont {Huang}}, \bibinfo {author} {\bibfnamefont
  {Z.}~\bibnamefont {Wang}}, \bibinfo {author} {\bibfnamefont {H.-Z.}\
  \bibnamefont {Lu}}, \bibinfo {author} {\bibfnamefont {D.}~\bibnamefont
  {Xing}}, \bibinfo {author} {\bibfnamefont {B.}~\bibnamefont {Wang}}, \bibinfo
  {author} {\bibfnamefont {X.}~\bibnamefont {Wan}},\ and\ \bibinfo {author}
  {\bibfnamefont {F.}~\bibnamefont {Miao}},\ }\bibfield  {title} {\bibinfo
  {title} {Gate-tunable negative longitudinal magnetoresistance in the
  predicted type-{{II Weyl}} semimetal {{WTe2}}},\ }\href
  {https://doi.org/10.1038/ncomms13142} {\bibfield  {journal} {\bibinfo
  {journal} {Nat Commun}\ }\textbf {\bibinfo {volume} {7}},\ \bibinfo {pages}
  {13142} (\bibinfo {year} {2016})}\BibitemShut {NoStop}%
\bibitem [{\citenamefont {Burkov}(2014)}]{burkov_anomalous_2014}%
  \BibitemOpen
  \bibfield  {author} {\bibinfo {author} {\bibfnamefont {A.~A.}\ \bibnamefont
  {Burkov}},\ }\bibfield  {title} {\bibinfo {title} {Anomalous {{Hall Effect}}
  in {{Weyl Metals}}},\ }\href {https://doi.org/10.1103/PhysRevLett.113.187202}
  {\bibfield  {journal} {\bibinfo  {journal} {Phys. Rev. Lett.}\ }\textbf
  {\bibinfo {volume} {113}},\ \bibinfo {pages} {187202} (\bibinfo {year}
  {2014})}\BibitemShut {NoStop}%
\bibitem [{\citenamefont {Steiner}\ \emph {et~al.}(2017)\citenamefont
  {Steiner}, \citenamefont {Andreev},\ and\ \citenamefont
  {Pesin}}]{steiner_anomalous_2017}%
  \BibitemOpen
  \bibfield  {author} {\bibinfo {author} {\bibfnamefont {J.~F.}\ \bibnamefont
  {Steiner}}, \bibinfo {author} {\bibfnamefont {A.~V.}\ \bibnamefont
  {Andreev}},\ and\ \bibinfo {author} {\bibfnamefont {D.~A.}\ \bibnamefont
  {Pesin}},\ }\bibfield  {title} {\bibinfo {title} {Anomalous {{Hall Effect}}
  in {{Type-I Weyl Metals}}},\ }\href
  {https://doi.org/10.1103/PhysRevLett.119.036601} {\bibfield  {journal}
  {\bibinfo  {journal} {Phys. Rev. Lett.}\ }\textbf {\bibinfo {volume} {119}},\
  \bibinfo {pages} {036601} (\bibinfo {year} {2017})}\BibitemShut {NoStop}%
\bibitem [{\citenamefont {Shekhar}\ \emph {et~al.}(2018)\citenamefont
  {Shekhar}, \citenamefont {Kumar}, \citenamefont {Grinenko}, \citenamefont
  {Singh}, \citenamefont {Sarkar}, \citenamefont {Luetkens}, \citenamefont
  {Wu}, \citenamefont {Zhang}, \citenamefont {Komarek}, \citenamefont
  {Kampert}, \citenamefont {Skourski}, \citenamefont {Wosnitza}, \citenamefont
  {Schnelle}, \citenamefont {McCollam}, \citenamefont {Zeitler}, \citenamefont
  {K{\"u}bler}, \citenamefont {Yan}, \citenamefont {Klauss}, \citenamefont
  {Parkin},\ and\ \citenamefont {Felser}}]{shekhar_anomalous_2018}%
  \BibitemOpen
  \bibfield  {author} {\bibinfo {author} {\bibfnamefont {C.}~\bibnamefont
  {Shekhar}}, \bibinfo {author} {\bibfnamefont {N.}~\bibnamefont {Kumar}},
  \bibinfo {author} {\bibfnamefont {V.}~\bibnamefont {Grinenko}}, \bibinfo
  {author} {\bibfnamefont {S.}~\bibnamefont {Singh}}, \bibinfo {author}
  {\bibfnamefont {R.}~\bibnamefont {Sarkar}}, \bibinfo {author} {\bibfnamefont
  {H.}~\bibnamefont {Luetkens}}, \bibinfo {author} {\bibfnamefont {S.-C.}\
  \bibnamefont {Wu}}, \bibinfo {author} {\bibfnamefont {Y.}~\bibnamefont
  {Zhang}}, \bibinfo {author} {\bibfnamefont {A.~C.}\ \bibnamefont {Komarek}},
  \bibinfo {author} {\bibfnamefont {E.}~\bibnamefont {Kampert}}, \bibinfo
  {author} {\bibfnamefont {Y.}~\bibnamefont {Skourski}}, \bibinfo {author}
  {\bibfnamefont {J.}~\bibnamefont {Wosnitza}}, \bibinfo {author}
  {\bibfnamefont {W.}~\bibnamefont {Schnelle}}, \bibinfo {author}
  {\bibfnamefont {A.}~\bibnamefont {McCollam}}, \bibinfo {author}
  {\bibfnamefont {U.}~\bibnamefont {Zeitler}}, \bibinfo {author} {\bibfnamefont
  {J.}~\bibnamefont {K{\"u}bler}}, \bibinfo {author} {\bibfnamefont
  {B.}~\bibnamefont {Yan}}, \bibinfo {author} {\bibfnamefont {H.-H.}\
  \bibnamefont {Klauss}}, \bibinfo {author} {\bibfnamefont {S.~S.~P.}\
  \bibnamefont {Parkin}},\ and\ \bibinfo {author} {\bibfnamefont
  {C.}~\bibnamefont {Felser}},\ }\bibfield  {title} {\bibinfo {title}
  {Anomalous {{Hall}} effect in {{Weyl}} semimetal half-{{Heusler}} compounds
  {{RPtBi}} ({{R}} = {{Gd}} and {{Nd}})},\ }\href
  {https://doi.org/10.1073/pnas.1810842115} {\bibfield  {journal} {\bibinfo
  {journal} {Proc. Natl. Acad. Sci. U.S.A.}\ }\textbf {\bibinfo {volume}
  {115}},\ \bibinfo {pages} {9140} (\bibinfo {year} {2018})}\BibitemShut
  {NoStop}%
\bibitem [{\citenamefont {Roy~Karmakar}\ \emph {et~al.}(2022)\citenamefont
  {Roy~Karmakar}, \citenamefont {Nandy}, \citenamefont {Taraphder},\ and\
  \citenamefont {Das}}]{roykarmakar_giant_2022}%
  \BibitemOpen
  \bibfield  {author} {\bibinfo {author} {\bibfnamefont {A.}~\bibnamefont
  {Roy~Karmakar}}, \bibinfo {author} {\bibfnamefont {S.}~\bibnamefont {Nandy}},
  \bibinfo {author} {\bibfnamefont {A.}~\bibnamefont {Taraphder}},\ and\
  \bibinfo {author} {\bibfnamefont {G.~P.}\ \bibnamefont {Das}},\ }\bibfield
  {title} {\bibinfo {title} {Giant anomalous thermal {{Hall}} effect in tilted
  type-{{I}} magnetic {{Weyl}} semimetal {{Co}} 3 {{Sn}} 2 {{S}} 2},\ }\href
  {https://doi.org/10.1103/PhysRevB.106.245133} {\bibfield  {journal} {\bibinfo
   {journal} {Phys. Rev. B}\ }\textbf {\bibinfo {volume} {106}},\ \bibinfo
  {pages} {245133} (\bibinfo {year} {2022})}\BibitemShut {NoStop}%
\bibitem [{\citenamefont {Gorbar}\ \emph {et~al.}(2017)\citenamefont {Gorbar},
  \citenamefont {Miransky}, \citenamefont {Shovkovy},\ and\ \citenamefont
  {Sukhachov}}]{gorbar_anomalous_2017}%
  \BibitemOpen
  \bibfield  {author} {\bibinfo {author} {\bibfnamefont {E.~V.}\ \bibnamefont
  {Gorbar}}, \bibinfo {author} {\bibfnamefont {V.~A.}\ \bibnamefont
  {Miransky}}, \bibinfo {author} {\bibfnamefont {I.~A.}\ \bibnamefont
  {Shovkovy}},\ and\ \bibinfo {author} {\bibfnamefont {P.~O.}\ \bibnamefont
  {Sukhachov}},\ }\bibfield  {title} {\bibinfo {title} {Anomalous
  thermoelectric phenomena in lattice models of multi-{{Weyl}} semimetals},\
  }\href {https://doi.org/10.1103/PhysRevB.96.155138} {\bibfield  {journal}
  {\bibinfo  {journal} {Phys. Rev. B}\ }\textbf {\bibinfo {volume} {96}},\
  \bibinfo {pages} {155138} (\bibinfo {year} {2017})}\BibitemShut {NoStop}%
\bibitem [{\citenamefont {Ferreiros}\ \emph {et~al.}(2017)\citenamefont
  {Ferreiros}, \citenamefont {Zyuzin},\ and\ \citenamefont
  {Bardarson}}]{ferreiros_anomalous_2017}%
  \BibitemOpen
  \bibfield  {author} {\bibinfo {author} {\bibfnamefont {Y.}~\bibnamefont
  {Ferreiros}}, \bibinfo {author} {\bibfnamefont {A.~A.}\ \bibnamefont
  {Zyuzin}},\ and\ \bibinfo {author} {\bibfnamefont {J.~H.}\ \bibnamefont
  {Bardarson}},\ }\bibfield  {title} {\bibinfo {title} {Anomalous {{Nernst}}
  and thermal {{Hall}} effects in tilted {{Weyl}} semimetals},\ }\href
  {https://doi.org/10.1103/PhysRevB.96.115202} {\bibfield  {journal} {\bibinfo
  {journal} {Phys. Rev. B}\ }\textbf {\bibinfo {volume} {96}},\ \bibinfo
  {pages} {115202} (\bibinfo {year} {2017})}\BibitemShut {NoStop}%
\bibitem [{\citenamefont {Watzman}\ \emph {et~al.}(2018)\citenamefont
  {Watzman}, \citenamefont {McCormick}, \citenamefont {Shekhar}, \citenamefont
  {Wu}, \citenamefont {Sun}, \citenamefont {Prakash}, \citenamefont {Felser},
  \citenamefont {Trivedi},\ and\ \citenamefont
  {Heremans}}]{watzman_dirac_2018}%
  \BibitemOpen
  \bibfield  {author} {\bibinfo {author} {\bibfnamefont {S.~J.}\ \bibnamefont
  {Watzman}}, \bibinfo {author} {\bibfnamefont {T.~M.}\ \bibnamefont
  {McCormick}}, \bibinfo {author} {\bibfnamefont {C.}~\bibnamefont {Shekhar}},
  \bibinfo {author} {\bibfnamefont {S.-C.}\ \bibnamefont {Wu}}, \bibinfo
  {author} {\bibfnamefont {Y.}~\bibnamefont {Sun}}, \bibinfo {author}
  {\bibfnamefont {A.}~\bibnamefont {Prakash}}, \bibinfo {author} {\bibfnamefont
  {C.}~\bibnamefont {Felser}}, \bibinfo {author} {\bibfnamefont
  {N.}~\bibnamefont {Trivedi}},\ and\ \bibinfo {author} {\bibfnamefont {J.~P.}\
  \bibnamefont {Heremans}},\ }\bibfield  {title} {\bibinfo {title} {Dirac
  dispersion generates unusually large {{Nernst}} effect in {{Weyl}}
  semimetals},\ }\href {https://doi.org/10.1103/PhysRevB.97.161404} {\bibfield
  {journal} {\bibinfo  {journal} {Phys. Rev. B}\ }\textbf {\bibinfo {volume}
  {97}},\ \bibinfo {pages} {161404} (\bibinfo {year} {2018})}\BibitemShut
  {NoStop}%
\bibitem [{\citenamefont {Lv}\ \emph {et~al.}(2015)\citenamefont {Lv},
  \citenamefont {Xu}, \citenamefont {Weng}, \citenamefont {Ma}, \citenamefont
  {Richard}, \citenamefont {Huang}, \citenamefont {Zhao}, \citenamefont {Chen},
  \citenamefont {Matt}, \citenamefont {Bisti}, \citenamefont {Strocov},
  \citenamefont {Mesot}, \citenamefont {Fang}, \citenamefont {Dai},
  \citenamefont {Qian}, \citenamefont {Shi},\ and\ \citenamefont
  {Ding}}]{lv_observation_2015}%
  \BibitemOpen
  \bibfield  {author} {\bibinfo {author} {\bibfnamefont {B.~Q.}\ \bibnamefont
  {Lv}}, \bibinfo {author} {\bibfnamefont {N.}~\bibnamefont {Xu}}, \bibinfo
  {author} {\bibfnamefont {H.~M.}\ \bibnamefont {Weng}}, \bibinfo {author}
  {\bibfnamefont {J.~Z.}\ \bibnamefont {Ma}}, \bibinfo {author} {\bibfnamefont
  {P.}~\bibnamefont {Richard}}, \bibinfo {author} {\bibfnamefont {X.~C.}\
  \bibnamefont {Huang}}, \bibinfo {author} {\bibfnamefont {L.~X.}\ \bibnamefont
  {Zhao}}, \bibinfo {author} {\bibfnamefont {G.~F.}\ \bibnamefont {Chen}},
  \bibinfo {author} {\bibfnamefont {C.~E.}\ \bibnamefont {Matt}}, \bibinfo
  {author} {\bibfnamefont {F.}~\bibnamefont {Bisti}}, \bibinfo {author}
  {\bibfnamefont {V.~N.}\ \bibnamefont {Strocov}}, \bibinfo {author}
  {\bibfnamefont {J.}~\bibnamefont {Mesot}}, \bibinfo {author} {\bibfnamefont
  {Z.}~\bibnamefont {Fang}}, \bibinfo {author} {\bibfnamefont {X.}~\bibnamefont
  {Dai}}, \bibinfo {author} {\bibfnamefont {T.}~\bibnamefont {Qian}}, \bibinfo
  {author} {\bibfnamefont {M.}~\bibnamefont {Shi}},\ and\ \bibinfo {author}
  {\bibfnamefont {H.}~\bibnamefont {Ding}},\ }\bibfield  {title} {\bibinfo
  {title} {Observation of {{Weyl}} nodes in {{TaAs}}},\ }\href
  {https://doi.org/10.1038/nphys3426} {\bibfield  {journal} {\bibinfo
  {journal} {Nature Phys}\ }\textbf {\bibinfo {volume} {11}},\ \bibinfo {pages}
  {724} (\bibinfo {year} {2015})}\BibitemShut {NoStop}%
\bibitem [{\citenamefont {Xu}\ \emph {et~al.}(2015)\citenamefont {Xu},
  \citenamefont {Belopolski}, \citenamefont {Sanchez}, \citenamefont {Zhang},
  \citenamefont {Chang}, \citenamefont {Guo}, \citenamefont {Bian},
  \citenamefont {Yuan}, \citenamefont {Lu}, \citenamefont {Chang},
  \citenamefont {Shibayev}, \citenamefont {Prokopovych}, \citenamefont
  {Alidoust}, \citenamefont {Zheng}, \citenamefont {Lee}, \citenamefont
  {Huang}, \citenamefont {Sankar}, \citenamefont {Chou}, \citenamefont {Hsu},
  \citenamefont {Jeng}, \citenamefont {Bansil}, \citenamefont {Neupert},
  \citenamefont {Strocov}, \citenamefont {Lin}, \citenamefont {Jia},\ and\
  \citenamefont {Hasan}}]{xu_experimental_2015}%
  \BibitemOpen
  \bibfield  {author} {\bibinfo {author} {\bibfnamefont {S.-Y.}\ \bibnamefont
  {Xu}}, \bibinfo {author} {\bibfnamefont {I.}~\bibnamefont {Belopolski}},
  \bibinfo {author} {\bibfnamefont {D.~S.}\ \bibnamefont {Sanchez}}, \bibinfo
  {author} {\bibfnamefont {C.}~\bibnamefont {Zhang}}, \bibinfo {author}
  {\bibfnamefont {G.}~\bibnamefont {Chang}}, \bibinfo {author} {\bibfnamefont
  {C.}~\bibnamefont {Guo}}, \bibinfo {author} {\bibfnamefont {G.}~\bibnamefont
  {Bian}}, \bibinfo {author} {\bibfnamefont {Z.}~\bibnamefont {Yuan}}, \bibinfo
  {author} {\bibfnamefont {H.}~\bibnamefont {Lu}}, \bibinfo {author}
  {\bibfnamefont {T.-R.}\ \bibnamefont {Chang}}, \bibinfo {author}
  {\bibfnamefont {P.~P.}\ \bibnamefont {Shibayev}}, \bibinfo {author}
  {\bibfnamefont {M.~L.}\ \bibnamefont {Prokopovych}}, \bibinfo {author}
  {\bibfnamefont {N.}~\bibnamefont {Alidoust}}, \bibinfo {author}
  {\bibfnamefont {H.}~\bibnamefont {Zheng}}, \bibinfo {author} {\bibfnamefont
  {C.-C.}\ \bibnamefont {Lee}}, \bibinfo {author} {\bibfnamefont {S.-M.}\
  \bibnamefont {Huang}}, \bibinfo {author} {\bibfnamefont {R.}~\bibnamefont
  {Sankar}}, \bibinfo {author} {\bibfnamefont {F.}~\bibnamefont {Chou}},
  \bibinfo {author} {\bibfnamefont {C.-H.}\ \bibnamefont {Hsu}}, \bibinfo
  {author} {\bibfnamefont {H.-T.}\ \bibnamefont {Jeng}}, \bibinfo {author}
  {\bibfnamefont {A.}~\bibnamefont {Bansil}}, \bibinfo {author} {\bibfnamefont
  {T.}~\bibnamefont {Neupert}}, \bibinfo {author} {\bibfnamefont {V.~N.}\
  \bibnamefont {Strocov}}, \bibinfo {author} {\bibfnamefont {H.}~\bibnamefont
  {Lin}}, \bibinfo {author} {\bibfnamefont {S.}~\bibnamefont {Jia}},\ and\
  \bibinfo {author} {\bibfnamefont {M.~Z.}\ \bibnamefont {Hasan}},\ }\bibfield
  {title} {\bibinfo {title} {Experimental discovery of a topological {{Weyl}}
  semimetal state in {{TaP}}},\ }\href {https://doi.org/10.1126/sciadv.1501092}
  {\bibfield  {journal} {\bibinfo  {journal} {Sci. Adv.}\ }\textbf {\bibinfo
  {volume} {1}},\ \bibinfo {pages} {e1501092} (\bibinfo {year}
  {2015})}\BibitemShut {NoStop}%
\bibitem [{\citenamefont {Weng}\ \emph {et~al.}(2015)\citenamefont {Weng},
  \citenamefont {Fang}, \citenamefont {Fang}, \citenamefont {Bernevig},\ and\
  \citenamefont {Dai}}]{weng_weyl_2015}%
  \BibitemOpen
  \bibfield  {author} {\bibinfo {author} {\bibfnamefont {H.}~\bibnamefont
  {Weng}}, \bibinfo {author} {\bibfnamefont {C.}~\bibnamefont {Fang}}, \bibinfo
  {author} {\bibfnamefont {Z.}~\bibnamefont {Fang}}, \bibinfo {author}
  {\bibfnamefont {B.~A.}\ \bibnamefont {Bernevig}},\ and\ \bibinfo {author}
  {\bibfnamefont {X.}~\bibnamefont {Dai}},\ }\bibfield  {title} {\bibinfo
  {title} {Weyl {{Semimetal Phase}} in {{Noncentrosymmetric Transition-Metal
  Monophosphides}}},\ }\href {https://doi.org/10.1103/PhysRevX.5.011029}
  {\bibfield  {journal} {\bibinfo  {journal} {Phys. Rev. X}\ }\textbf {\bibinfo
  {volume} {5}},\ \bibinfo {pages} {011029} (\bibinfo {year}
  {2015})}\BibitemShut {NoStop}%
\bibitem [{\citenamefont {Dey}\ \emph {et~al.}(2020)\citenamefont {Dey},
  \citenamefont {Nandy},\ and\ \citenamefont {Taraphder}}]{dey_dynamic_2020}%
  \BibitemOpen
  \bibfield  {author} {\bibinfo {author} {\bibfnamefont {U.}~\bibnamefont
  {Dey}}, \bibinfo {author} {\bibfnamefont {S.}~\bibnamefont {Nandy}},\ and\
  \bibinfo {author} {\bibfnamefont {A.}~\bibnamefont {Taraphder}},\ }\bibfield
  {title} {\bibinfo {title} {Dynamic chiral magnetic effect and anisotropic
  natural optical activity of tilted {{Weyl}} semimetals},\ }\href
  {https://doi.org/10.1038/s41598-020-59385-6} {\bibfield  {journal} {\bibinfo
  {journal} {Sci Rep}\ }\textbf {\bibinfo {volume} {10}},\ \bibinfo {pages}
  {2699} (\bibinfo {year} {2020})}\BibitemShut {NoStop}%
\bibitem [{\citenamefont {Sodemann}\ and\ \citenamefont
  {Fu}(2015)}]{sodemann_quantum_2015}%
  \BibitemOpen
  \bibfield  {author} {\bibinfo {author} {\bibfnamefont {I.}~\bibnamefont
  {Sodemann}}\ and\ \bibinfo {author} {\bibfnamefont {L.}~\bibnamefont {Fu}},\
  }\bibfield  {title} {\bibinfo {title} {Quantum {{Nonlinear Hall Effect
  Induced}} by {{Berry Curvature Dipole}} in {{Time-Reversal Invariant
  Materials}}},\ }\href {https://doi.org/10.1103/PhysRevLett.115.216806}
  {\bibfield  {journal} {\bibinfo  {journal} {Phys. Rev. Lett.}\ }\textbf
  {\bibinfo {volume} {115}},\ \bibinfo {pages} {216806} (\bibinfo {year}
  {2015})}\BibitemShut {NoStop}%
\bibitem [{\citenamefont {Kumar}\ \emph {et~al.}(2021)\citenamefont {Kumar},
  \citenamefont {Hsu}, \citenamefont {Sharma}, \citenamefont {Chang},
  \citenamefont {Yu}, \citenamefont {Wang}, \citenamefont {Eda}, \citenamefont
  {Liang},\ and\ \citenamefont {Yang}}]{kumar_roomtemperature_2021}%
  \BibitemOpen
  \bibfield  {author} {\bibinfo {author} {\bibfnamefont {D.}~\bibnamefont
  {Kumar}}, \bibinfo {author} {\bibfnamefont {C.-H.}\ \bibnamefont {Hsu}},
  \bibinfo {author} {\bibfnamefont {R.}~\bibnamefont {Sharma}}, \bibinfo
  {author} {\bibfnamefont {T.-R.}\ \bibnamefont {Chang}}, \bibinfo {author}
  {\bibfnamefont {P.}~\bibnamefont {Yu}}, \bibinfo {author} {\bibfnamefont
  {J.}~\bibnamefont {Wang}}, \bibinfo {author} {\bibfnamefont {G.}~\bibnamefont
  {Eda}}, \bibinfo {author} {\bibfnamefont {G.}~\bibnamefont {Liang}},\ and\
  \bibinfo {author} {\bibfnamefont {H.}~\bibnamefont {Yang}},\ }\bibfield
  {title} {\bibinfo {title} {Room-temperature nonlinear {{Hall}} effect and
  wireless radiofrequency rectification in {{Weyl}} semimetal {{TaIrTe4}}},\
  }\href {https://doi.org/10.1038/s41565-020-00839-3} {\bibfield  {journal}
  {\bibinfo  {journal} {Nat. Nanotechnol.}\ }\textbf {\bibinfo {volume} {16}},\
  \bibinfo {pages} {421} (\bibinfo {year} {2021})}\BibitemShut {NoStop}%
\bibitem [{\citenamefont {Zeng}\ \emph {et~al.}(2021)\citenamefont {Zeng},
  \citenamefont {Nandy},\ and\ \citenamefont {Tewari}}]{zeng_nonlinear_2021}%
  \BibitemOpen
  \bibfield  {author} {\bibinfo {author} {\bibfnamefont {C.}~\bibnamefont
  {Zeng}}, \bibinfo {author} {\bibfnamefont {S.}~\bibnamefont {Nandy}},\ and\
  \bibinfo {author} {\bibfnamefont {S.}~\bibnamefont {Tewari}},\ }\bibfield
  {title} {\bibinfo {title} {Nonlinear transport in {{Weyl}} semimetals induced
  by {{Berry}} curvature dipole},\ }\href
  {https://doi.org/10.1103/PhysRevB.103.245119} {\bibfield  {journal} {\bibinfo
   {journal} {Phys. Rev. B}\ }\textbf {\bibinfo {volume} {103}},\ \bibinfo
  {pages} {245119} (\bibinfo {year} {2021})}\BibitemShut {NoStop}%
\bibitem [{\citenamefont {Lv}\ \emph {et~al.}(2021)\citenamefont {Lv},
  \citenamefont {Xu}, \citenamefont {Han}, \citenamefont {Zhang}, \citenamefont
  {Han}, \citenamefont {Zhou}, \citenamefont {Yao}, \citenamefont {Liu},
  \citenamefont {Lu}, \citenamefont {Weng}, \citenamefont {Xie}, \citenamefont
  {Chen}, \citenamefont {Hu}, \citenamefont {Chen},\ and\ \citenamefont
  {Zhu}}]{lv_highharmonic_2021}%
  \BibitemOpen
  \bibfield  {author} {\bibinfo {author} {\bibfnamefont {Y.-Y.}\ \bibnamefont
  {Lv}}, \bibinfo {author} {\bibfnamefont {J.}~\bibnamefont {Xu}}, \bibinfo
  {author} {\bibfnamefont {S.}~\bibnamefont {Han}}, \bibinfo {author}
  {\bibfnamefont {C.}~\bibnamefont {Zhang}}, \bibinfo {author} {\bibfnamefont
  {Y.}~\bibnamefont {Han}}, \bibinfo {author} {\bibfnamefont {J.}~\bibnamefont
  {Zhou}}, \bibinfo {author} {\bibfnamefont {S.-H.}\ \bibnamefont {Yao}},
  \bibinfo {author} {\bibfnamefont {X.-P.}\ \bibnamefont {Liu}}, \bibinfo
  {author} {\bibfnamefont {M.-H.}\ \bibnamefont {Lu}}, \bibinfo {author}
  {\bibfnamefont {H.}~\bibnamefont {Weng}}, \bibinfo {author} {\bibfnamefont
  {Z.}~\bibnamefont {Xie}}, \bibinfo {author} {\bibfnamefont {Y.~B.}\
  \bibnamefont {Chen}}, \bibinfo {author} {\bibfnamefont {J.}~\bibnamefont
  {Hu}}, \bibinfo {author} {\bibfnamefont {Y.-F.}\ \bibnamefont {Chen}},\ and\
  \bibinfo {author} {\bibfnamefont {S.}~\bibnamefont {Zhu}},\ }\bibfield
  {title} {\bibinfo {title} {High-harmonic generation in {{Weyl}} semimetal
  {$\beta$}-{{WP2}} crystals},\ }\href
  {https://doi.org/10.1038/s41467-021-26766-y} {\bibfield  {journal} {\bibinfo
  {journal} {Nat Commun}\ }\textbf {\bibinfo {volume} {12}},\ \bibinfo {pages}
  {6437} (\bibinfo {year} {2021})}\BibitemShut {NoStop}%
\bibitem [{\citenamefont {Lu}\ \emph {et~al.}(2022)\citenamefont {Lu},
  \citenamefont {Sayyad}, \citenamefont {{S{\'a}nchez-Mart{\'i}nez}},
  \citenamefont {Manna}, \citenamefont {Felser}, \citenamefont {Grushin},\ and\
  \citenamefont {Torchinsky}}]{lu_secondharmonic_2022}%
  \BibitemOpen
  \bibfield  {author} {\bibinfo {author} {\bibfnamefont {B.}~\bibnamefont
  {Lu}}, \bibinfo {author} {\bibfnamefont {S.}~\bibnamefont {Sayyad}}, \bibinfo
  {author} {\bibfnamefont {M.~{\'A}.}\ \bibnamefont
  {{S{\'a}nchez-Mart{\'i}nez}}}, \bibinfo {author} {\bibfnamefont
  {K.}~\bibnamefont {Manna}}, \bibinfo {author} {\bibfnamefont
  {C.}~\bibnamefont {Felser}}, \bibinfo {author} {\bibfnamefont {A.~G.}\
  \bibnamefont {Grushin}},\ and\ \bibinfo {author} {\bibfnamefont {D.~H.}\
  \bibnamefont {Torchinsky}},\ }\bibfield  {title} {\bibinfo {title}
  {Second-harmonic generation in the topological multifold semimetal
  {{RhSi}}},\ }\href {https://doi.org/10.1103/PhysRevResearch.4.L022022}
  {\bibfield  {journal} {\bibinfo  {journal} {Phys. Rev. Research}\ }\textbf
  {\bibinfo {volume} {4}},\ \bibinfo {pages} {L022022} (\bibinfo {year}
  {2022})}\BibitemShut {NoStop}%
\bibitem [{\citenamefont {Li}\ \emph {et~al.}(2018)\citenamefont {Li},
  \citenamefont {Jin}, \citenamefont {Tohyama}, \citenamefont {Iitaka},
  \citenamefont {Zhang},\ and\ \citenamefont {Su}}]{li_second_2018}%
  \BibitemOpen
  \bibfield  {author} {\bibinfo {author} {\bibfnamefont {Z.}~\bibnamefont
  {Li}}, \bibinfo {author} {\bibfnamefont {Y.-Q.}\ \bibnamefont {Jin}},
  \bibinfo {author} {\bibfnamefont {T.}~\bibnamefont {Tohyama}}, \bibinfo
  {author} {\bibfnamefont {T.}~\bibnamefont {Iitaka}}, \bibinfo {author}
  {\bibfnamefont {J.-X.}\ \bibnamefont {Zhang}},\ and\ \bibinfo {author}
  {\bibfnamefont {H.}~\bibnamefont {Su}},\ }\bibfield  {title} {\bibinfo
  {title} {Second harmonic generation in the {{Weyl}} semimetal {{TaAs}} from a
  quantum kinetic equation},\ }\href
  {https://doi.org/10.1103/PhysRevB.97.085201} {\bibfield  {journal} {\bibinfo
  {journal} {Phys. Rev. B}\ }\textbf {\bibinfo {volume} {97}},\ \bibinfo
  {pages} {085201} (\bibinfo {year} {2018})}\BibitemShut {NoStop}%
\bibitem [{\citenamefont {Ahn}\ \emph {et~al.}(2020)\citenamefont {Ahn},
  \citenamefont {Guo},\ and\ \citenamefont {Nagaosa}}]{ahn_lowfrequency_2020}%
  \BibitemOpen
  \bibfield  {author} {\bibinfo {author} {\bibfnamefont {J.}~\bibnamefont
  {Ahn}}, \bibinfo {author} {\bibfnamefont {G.-Y.}\ \bibnamefont {Guo}},\ and\
  \bibinfo {author} {\bibfnamefont {N.}~\bibnamefont {Nagaosa}},\ }\bibfield
  {title} {\bibinfo {title} {Low-{{Frequency Divergence}} and {{Quantum
  Geometry}} of the {{Bulk Photovoltaic Effect}} in {{Topological
  Semimetals}}},\ }\href {https://doi.org/10.1103/PhysRevX.10.041041}
  {\bibfield  {journal} {\bibinfo  {journal} {Phys. Rev. X}\ }\textbf {\bibinfo
  {volume} {10}},\ \bibinfo {pages} {041041} (\bibinfo {year}
  {2020})}\BibitemShut {NoStop}%
\bibitem [{\citenamefont {Osterhoudt}\ \emph {et~al.}(2019)\citenamefont
  {Osterhoudt}, \citenamefont {Diebel}, \citenamefont {Gray}, \citenamefont
  {Yang}, \citenamefont {Stanco}, \citenamefont {Huang}, \citenamefont {Shen},
  \citenamefont {Ni}, \citenamefont {Moll}, \citenamefont {Ran},\ and\
  \citenamefont {Burch}}]{osterhoudt_colossal_2019}%
  \BibitemOpen
  \bibfield  {author} {\bibinfo {author} {\bibfnamefont {G.~B.}\ \bibnamefont
  {Osterhoudt}}, \bibinfo {author} {\bibfnamefont {L.~K.}\ \bibnamefont
  {Diebel}}, \bibinfo {author} {\bibfnamefont {M.~J.}\ \bibnamefont {Gray}},
  \bibinfo {author} {\bibfnamefont {X.}~\bibnamefont {Yang}}, \bibinfo {author}
  {\bibfnamefont {J.}~\bibnamefont {Stanco}}, \bibinfo {author} {\bibfnamefont
  {X.}~\bibnamefont {Huang}}, \bibinfo {author} {\bibfnamefont
  {B.}~\bibnamefont {Shen}}, \bibinfo {author} {\bibfnamefont {N.}~\bibnamefont
  {Ni}}, \bibinfo {author} {\bibfnamefont {P.~J.~W.}\ \bibnamefont {Moll}},
  \bibinfo {author} {\bibfnamefont {Y.}~\bibnamefont {Ran}},\ and\ \bibinfo
  {author} {\bibfnamefont {K.~S.}\ \bibnamefont {Burch}},\ }\bibfield  {title}
  {\bibinfo {title} {Colossal mid-infrared bulk photovoltaic effect in a
  type-{{I Weyl}} semimetal},\ }\href
  {https://doi.org/10.1038/s41563-019-0297-4} {\bibfield  {journal} {\bibinfo
  {journal} {Nat. Mater.}\ }\textbf {\bibinfo {volume} {18}},\ \bibinfo {pages}
  {471} (\bibinfo {year} {2019})}\BibitemShut {NoStop}%
\bibitem [{\citenamefont {Tiwari}\ \emph {et~al.}(2020)\citenamefont {Tiwari},
  \citenamefont {Birajdar},\ and\ \citenamefont
  {Ghosh}}]{tiwari_firstprinciples_2020}%
  \BibitemOpen
  \bibfield  {author} {\bibinfo {author} {\bibfnamefont {R.~P.}\ \bibnamefont
  {Tiwari}}, \bibinfo {author} {\bibfnamefont {B.}~\bibnamefont {Birajdar}},\
  and\ \bibinfo {author} {\bibfnamefont {R.~K.}\ \bibnamefont {Ghosh}},\
  }\bibfield  {title} {\bibinfo {title} {First-principles calculation of shift
  current bulk photovoltaic effect in two-dimensional {$\alpha$}- {{I}} n 2
  {{S}} e 3},\ }\href {https://doi.org/10.1103/PhysRevB.101.235448} {\bibfield
  {journal} {\bibinfo  {journal} {Phys. Rev. B}\ }\textbf {\bibinfo {volume}
  {101}},\ \bibinfo {pages} {235448} (\bibinfo {year} {2020})}\BibitemShut
  {NoStop}%
\bibitem [{\citenamefont {De~Juan}\ \emph {et~al.}(2020)\citenamefont
  {De~Juan}, \citenamefont {Zhang}, \citenamefont {Morimoto}, \citenamefont
  {Sun}, \citenamefont {Moore},\ and\ \citenamefont
  {Grushin}}]{dejuan_difference_2020}%
  \BibitemOpen
  \bibfield  {author} {\bibinfo {author} {\bibfnamefont {F.}~\bibnamefont
  {De~Juan}}, \bibinfo {author} {\bibfnamefont {Y.}~\bibnamefont {Zhang}},
  \bibinfo {author} {\bibfnamefont {T.}~\bibnamefont {Morimoto}}, \bibinfo
  {author} {\bibfnamefont {Y.}~\bibnamefont {Sun}}, \bibinfo {author}
  {\bibfnamefont {J.~E.}\ \bibnamefont {Moore}},\ and\ \bibinfo {author}
  {\bibfnamefont {A.~G.}\ \bibnamefont {Grushin}},\ }\bibfield  {title}
  {\bibinfo {title} {Difference frequency generation in topological
  semimetals},\ }\href {https://doi.org/10.1103/PhysRevResearch.2.012017}
  {\bibfield  {journal} {\bibinfo  {journal} {Phys. Rev. Research}\ }\textbf
  {\bibinfo {volume} {2}},\ \bibinfo {pages} {012017} (\bibinfo {year}
  {2020})}\BibitemShut {NoStop}%
\bibitem [{\citenamefont {Flicker}\ \emph {et~al.}(2018)\citenamefont
  {Flicker}, \citenamefont {De~Juan}, \citenamefont {Bradlyn}, \citenamefont
  {Morimoto}, \citenamefont {Vergniory},\ and\ \citenamefont
  {Grushin}}]{flicker_chiral_2018}%
  \BibitemOpen
  \bibfield  {author} {\bibinfo {author} {\bibfnamefont {F.}~\bibnamefont
  {Flicker}}, \bibinfo {author} {\bibfnamefont {F.}~\bibnamefont {De~Juan}},
  \bibinfo {author} {\bibfnamefont {B.}~\bibnamefont {Bradlyn}}, \bibinfo
  {author} {\bibfnamefont {T.}~\bibnamefont {Morimoto}}, \bibinfo {author}
  {\bibfnamefont {M.~G.}\ \bibnamefont {Vergniory}},\ and\ \bibinfo {author}
  {\bibfnamefont {A.~G.}\ \bibnamefont {Grushin}},\ }\bibfield  {title}
  {\bibinfo {title} {Chiral optical response of multifold fermions},\ }\href
  {https://doi.org/10.1103/PhysRevB.98.155145} {\bibfield  {journal} {\bibinfo
  {journal} {Phys. Rev. B}\ }\textbf {\bibinfo {volume} {98}},\ \bibinfo
  {pages} {155145} (\bibinfo {year} {2018})}\BibitemShut {NoStop}%
\bibitem [{\citenamefont {Sadhukhan}\ and\ \citenamefont
  {Nag}(2021{\natexlab{a}})}]{sadhukhan_electronic_2021}%
  \BibitemOpen
  \bibfield  {author} {\bibinfo {author} {\bibfnamefont {B.}~\bibnamefont
  {Sadhukhan}}\ and\ \bibinfo {author} {\bibfnamefont {T.}~\bibnamefont
  {Nag}},\ }\bibfield  {title} {\bibinfo {title} {Electronic structure and
  unconventional nonlinear response in double {{Weyl}} semimetal {{Sr Si}} 2},\
  }\href {https://doi.org/10.1103/PhysRevB.104.245122} {\bibfield  {journal}
  {\bibinfo  {journal} {Phys. Rev. B}\ }\textbf {\bibinfo {volume} {104}},\
  \bibinfo {pages} {245122} (\bibinfo {year} {2021}{\natexlab{a}})}\BibitemShut
  {NoStop}%
\bibitem [{\citenamefont {Cook}\ \emph {et~al.}(2017)\citenamefont {Cook},
  \citenamefont {M.~Fregoso}, \citenamefont {De~Juan}, \citenamefont {Coh},\
  and\ \citenamefont {Moore}}]{cook_design_2017}%
  \BibitemOpen
  \bibfield  {author} {\bibinfo {author} {\bibfnamefont {A.~M.}\ \bibnamefont
  {Cook}}, \bibinfo {author} {\bibfnamefont {B.}~\bibnamefont {M.~Fregoso}},
  \bibinfo {author} {\bibfnamefont {F.}~\bibnamefont {De~Juan}}, \bibinfo
  {author} {\bibfnamefont {S.}~\bibnamefont {Coh}},\ and\ \bibinfo {author}
  {\bibfnamefont {J.~E.}\ \bibnamefont {Moore}},\ }\bibfield  {title} {\bibinfo
  {title} {Design principles for shift current photovoltaics},\ }\href
  {https://doi.org/10.1038/ncomms14176} {\bibfield  {journal} {\bibinfo
  {journal} {Nat Commun}\ }\textbf {\bibinfo {volume} {8}},\ \bibinfo {pages}
  {14176} (\bibinfo {year} {2017})}\BibitemShut {NoStop}%
\bibitem [{\citenamefont {Tan}\ \emph {et~al.}(2016)\citenamefont {Tan},
  \citenamefont {Zheng}, \citenamefont {Young}, \citenamefont {Wang},
  \citenamefont {Liu},\ and\ \citenamefont {Rappe}}]{tan_shift_2016}%
  \BibitemOpen
  \bibfield  {author} {\bibinfo {author} {\bibfnamefont {L.~Z.}\ \bibnamefont
  {Tan}}, \bibinfo {author} {\bibfnamefont {F.}~\bibnamefont {Zheng}}, \bibinfo
  {author} {\bibfnamefont {S.~M.}\ \bibnamefont {Young}}, \bibinfo {author}
  {\bibfnamefont {F.}~\bibnamefont {Wang}}, \bibinfo {author} {\bibfnamefont
  {S.}~\bibnamefont {Liu}},\ and\ \bibinfo {author} {\bibfnamefont {A.~M.}\
  \bibnamefont {Rappe}},\ }\bibfield  {title} {\bibinfo {title} {Shift current
  bulk photovoltaic effect in polar materials---hybrid and oxide perovskites
  and beyond},\ }\href {https://doi.org/10.1038/npjcompumats.2016.26}
  {\bibfield  {journal} {\bibinfo  {journal} {npj Comput Mater}\ }\textbf
  {\bibinfo {volume} {2}},\ \bibinfo {pages} {16026} (\bibinfo {year}
  {2016})}\BibitemShut {NoStop}%
\bibitem [{\citenamefont {Belinicher}\ and\ \citenamefont
  {Sturman}(1980)}]{belinicher_photogalvanic_1980}%
  \BibitemOpen
  \bibfield  {author} {\bibinfo {author} {\bibfnamefont {V.~I.}\ \bibnamefont
  {Belinicher}}\ and\ \bibinfo {author} {\bibfnamefont {B.~I.}\ \bibnamefont
  {Sturman}},\ }\bibfield  {title} {\bibinfo {title} {The photogalvanic effect
  in media lacking a center of symmetry},\ }\href
  {https://doi.org/10.1070/PU1980v023n03ABEH004703} {\bibfield  {journal}
  {\bibinfo  {journal} {Sov. Phys. Usp.}\ }\textbf {\bibinfo {volume} {23}},\
  \bibinfo {pages} {199} (\bibinfo {year} {1980})}\BibitemShut {NoStop}%
\bibitem [{\citenamefont {Asnin}\ \emph {et~al.}(1979)\citenamefont {Asnin},
  \citenamefont {Bakun}, \citenamefont {Danishevskii}, \citenamefont
  {Ivchenko}, \citenamefont {Pikus},\ and\ \citenamefont
  {Rogachev}}]{asnin_circular_1979}%
  \BibitemOpen
  \bibfield  {author} {\bibinfo {author} {\bibfnamefont {V.}~\bibnamefont
  {Asnin}}, \bibinfo {author} {\bibfnamefont {A.}~\bibnamefont {Bakun}},
  \bibinfo {author} {\bibfnamefont {A.}~\bibnamefont {Danishevskii}}, \bibinfo
  {author} {\bibfnamefont {E.}~\bibnamefont {Ivchenko}}, \bibinfo {author}
  {\bibfnamefont {G.}~\bibnamefont {Pikus}},\ and\ \bibinfo {author}
  {\bibfnamefont {A.}~\bibnamefont {Rogachev}},\ }\bibfield  {title} {\bibinfo
  {title} {``{{Circular}}'' photogalvanic effect in optically active
  crystals},\ }\href {https://doi.org/10.1016/0038-1098(79)91137-2} {\bibfield
  {journal} {\bibinfo  {journal} {Solid State Communications}\ }\textbf
  {\bibinfo {volume} {30}},\ \bibinfo {pages} {565} (\bibinfo {year}
  {1979})}\BibitemShut {NoStop}%
\bibitem [{\citenamefont {Pikus}\ and\ \citenamefont
  {Ivchenko}(1980)}]{pikus_photogalvanic_1980}%
  \BibitemOpen
  \bibfield  {author} {\bibinfo {author} {\bibfnamefont {G.~E.}\ \bibnamefont
  {Pikus}}\ and\ \bibinfo {author} {\bibfnamefont {E.~L.}\ \bibnamefont
  {Ivchenko}},\ }\bibfield  {title} {\bibinfo {title} {Photogalvanic effects in
  noncentrosymmetric semiconductors},\ }in\ \href
  {https://doi.org/10.1007/3-540-10261-2_54} {\emph {\bibinfo {booktitle}
  {Narrow {{Gap Semiconductors Physics}} and {{Applications}}}}},\ Vol.\
  \bibinfo {volume} {133},\ \bibinfo {editor} {edited by\ \bibinfo {editor}
  {\bibfnamefont {J.}~\bibnamefont {Ehlers}}, \bibinfo {editor} {\bibfnamefont
  {K.}~\bibnamefont {Hepp}}, \bibinfo {editor} {\bibfnamefont {R.}~\bibnamefont
  {Kippenhahn}}, \bibinfo {editor} {\bibfnamefont {H.~A.}\ \bibnamefont
  {Weidenm{\"u}ller}}, \bibinfo {editor} {\bibfnamefont {J.}~\bibnamefont
  {Zittartz}}, \bibinfo {editor} {\bibfnamefont {W.}~\bibnamefont
  {Beiglb{\"o}ck}},\ and\ \bibinfo {editor} {\bibfnamefont {W.}~\bibnamefont
  {Zawadzki}}}\ (\bibinfo  {publisher} {Springer Berlin Heidelberg},\ \bibinfo
  {address} {Berlin, Heidelberg},\ \bibinfo {year} {1980})\ pp.\ \bibinfo
  {pages} {388--406}\BibitemShut {NoStop}%
\bibitem [{\citenamefont {De~Juan}\ \emph {et~al.}(2017)\citenamefont
  {De~Juan}, \citenamefont {Grushin}, \citenamefont {Morimoto},\ and\
  \citenamefont {Moore}}]{dejuan_quantized_2017}%
  \BibitemOpen
  \bibfield  {author} {\bibinfo {author} {\bibfnamefont {F.}~\bibnamefont
  {De~Juan}}, \bibinfo {author} {\bibfnamefont {A.~G.}\ \bibnamefont
  {Grushin}}, \bibinfo {author} {\bibfnamefont {T.}~\bibnamefont {Morimoto}},\
  and\ \bibinfo {author} {\bibfnamefont {J.~E.}\ \bibnamefont {Moore}},\
  }\bibfield  {title} {\bibinfo {title} {Quantized circular photogalvanic
  effect in {{Weyl}} semimetals},\ }\href {https://doi.org/10.1038/ncomms15995}
  {\bibfield  {journal} {\bibinfo  {journal} {Nat Commun}\ }\textbf {\bibinfo
  {volume} {8}},\ \bibinfo {pages} {15995} (\bibinfo {year}
  {2017})}\BibitemShut {NoStop}%
\bibitem [{\citenamefont {Zhang}\ \emph {et~al.}(2018)\citenamefont {Zhang},
  \citenamefont {Ishizuka}, \citenamefont {Van Den~Brink}, \citenamefont
  {Felser}, \citenamefont {Yan},\ and\ \citenamefont
  {Nagaosa}}]{zhang_photogalvanic_2018}%
  \BibitemOpen
  \bibfield  {author} {\bibinfo {author} {\bibfnamefont {Y.}~\bibnamefont
  {Zhang}}, \bibinfo {author} {\bibfnamefont {H.}~\bibnamefont {Ishizuka}},
  \bibinfo {author} {\bibfnamefont {J.}~\bibnamefont {Van Den~Brink}}, \bibinfo
  {author} {\bibfnamefont {C.}~\bibnamefont {Felser}}, \bibinfo {author}
  {\bibfnamefont {B.}~\bibnamefont {Yan}},\ and\ \bibinfo {author}
  {\bibfnamefont {N.}~\bibnamefont {Nagaosa}},\ }\bibfield  {title} {\bibinfo
  {title} {Photogalvanic effect in {{Weyl}} semimetals from first principles},\
  }\href {https://doi.org/10.1103/PhysRevB.97.241118} {\bibfield  {journal}
  {\bibinfo  {journal} {Phys. Rev. B}\ }\textbf {\bibinfo {volume} {97}},\
  \bibinfo {pages} {241118} (\bibinfo {year} {2018})}\BibitemShut {NoStop}%
\bibitem [{\citenamefont {Wu}\ \emph {et~al.}(2017)\citenamefont {Wu},
  \citenamefont {Patankar}, \citenamefont {Morimoto}, \citenamefont {Nair},
  \citenamefont {Thewalt}, \citenamefont {Little}, \citenamefont {Analytis},
  \citenamefont {Moore},\ and\ \citenamefont {Orenstein}}]{wu_giant_2017}%
  \BibitemOpen
  \bibfield  {author} {\bibinfo {author} {\bibfnamefont {L.}~\bibnamefont
  {Wu}}, \bibinfo {author} {\bibfnamefont {S.}~\bibnamefont {Patankar}},
  \bibinfo {author} {\bibfnamefont {T.}~\bibnamefont {Morimoto}}, \bibinfo
  {author} {\bibfnamefont {N.~L.}\ \bibnamefont {Nair}}, \bibinfo {author}
  {\bibfnamefont {E.}~\bibnamefont {Thewalt}}, \bibinfo {author} {\bibfnamefont
  {A.}~\bibnamefont {Little}}, \bibinfo {author} {\bibfnamefont {J.~G.}\
  \bibnamefont {Analytis}}, \bibinfo {author} {\bibfnamefont {J.~E.}\
  \bibnamefont {Moore}},\ and\ \bibinfo {author} {\bibfnamefont
  {J.}~\bibnamefont {Orenstein}},\ }\bibfield  {title} {\bibinfo {title} {Giant
  anisotropic nonlinear optical response in transition metal monopnictide
  {{Weyl}} semimetals},\ }\href {https://doi.org/10.1038/nphys3969} {\bibfield
  {journal} {\bibinfo  {journal} {Nature Phys}\ }\textbf {\bibinfo {volume}
  {13}},\ \bibinfo {pages} {350} (\bibinfo {year} {2017})}\BibitemShut
  {NoStop}%
\bibitem [{\citenamefont {Nag}\ and\ \citenamefont
  {Kennes}(2022)}]{nag_distinct_2022}%
  \BibitemOpen
  \bibfield  {author} {\bibinfo {author} {\bibfnamefont {T.}~\bibnamefont
  {Nag}}\ and\ \bibinfo {author} {\bibfnamefont {D.~M.}\ \bibnamefont
  {Kennes}},\ }\bibfield  {title} {\bibinfo {title} {Distinct signatures of
  particle-hole symmetry breaking in transport coefficients for generic
  multi-{{Weyl}} semimetals},\ }\href
  {https://doi.org/10.1103/PhysRevB.105.214307} {\bibfield  {journal} {\bibinfo
   {journal} {Phys. Rev. B}\ }\textbf {\bibinfo {volume} {105}},\ \bibinfo
  {pages} {214307} (\bibinfo {year} {2022})}\BibitemShut {NoStop}%
\bibitem [{\citenamefont {Sessi}\ \emph {et~al.}(2020)\citenamefont {Sessi},
  \citenamefont {Fan}, \citenamefont {K{\"u}ster}, \citenamefont {Manna},
  \citenamefont {Schr{\"o}ter}, \citenamefont {Ji}, \citenamefont {Stolz},
  \citenamefont {Krieger}, \citenamefont {Pei}, \citenamefont {Kim},
  \citenamefont {Dudin}, \citenamefont {Cacho}, \citenamefont {Widmer},
  \citenamefont {Borrmann}, \citenamefont {Shi}, \citenamefont {Chang},
  \citenamefont {Sun}, \citenamefont {Felser},\ and\ \citenamefont
  {Parkin}}]{sessi_handednessdependent_2020}%
  \BibitemOpen
  \bibfield  {author} {\bibinfo {author} {\bibfnamefont {P.}~\bibnamefont
  {Sessi}}, \bibinfo {author} {\bibfnamefont {F.-R.}\ \bibnamefont {Fan}},
  \bibinfo {author} {\bibfnamefont {F.}~\bibnamefont {K{\"u}ster}}, \bibinfo
  {author} {\bibfnamefont {K.}~\bibnamefont {Manna}}, \bibinfo {author}
  {\bibfnamefont {N.~B.~M.}\ \bibnamefont {Schr{\"o}ter}}, \bibinfo {author}
  {\bibfnamefont {J.-R.}\ \bibnamefont {Ji}}, \bibinfo {author} {\bibfnamefont
  {S.}~\bibnamefont {Stolz}}, \bibinfo {author} {\bibfnamefont {J.~A.}\
  \bibnamefont {Krieger}}, \bibinfo {author} {\bibfnamefont {D.}~\bibnamefont
  {Pei}}, \bibinfo {author} {\bibfnamefont {T.~K.}\ \bibnamefont {Kim}},
  \bibinfo {author} {\bibfnamefont {P.}~\bibnamefont {Dudin}}, \bibinfo
  {author} {\bibfnamefont {C.}~\bibnamefont {Cacho}}, \bibinfo {author}
  {\bibfnamefont {R.}~\bibnamefont {Widmer}}, \bibinfo {author} {\bibfnamefont
  {H.}~\bibnamefont {Borrmann}}, \bibinfo {author} {\bibfnamefont
  {W.}~\bibnamefont {Shi}}, \bibinfo {author} {\bibfnamefont {K.}~\bibnamefont
  {Chang}}, \bibinfo {author} {\bibfnamefont {Y.}~\bibnamefont {Sun}}, \bibinfo
  {author} {\bibfnamefont {C.}~\bibnamefont {Felser}},\ and\ \bibinfo {author}
  {\bibfnamefont {S.~S.~P.}\ \bibnamefont {Parkin}},\ }\bibfield  {title}
  {\bibinfo {title} {Handedness-dependent quasiparticle interference in the two
  enantiomers of the topological chiral semimetal {{PdGa}}},\ }\href
  {https://doi.org/10.1038/s41467-020-17261-x} {\bibfield  {journal} {\bibinfo
  {journal} {Nat Commun}\ }\textbf {\bibinfo {volume} {11}},\ \bibinfo {pages}
  {3507} (\bibinfo {year} {2020})}\BibitemShut {NoStop}%
\bibitem [{\citenamefont {{Iba{\~n}ez-Azpiroz}}\ \emph
  {et~al.}(2018)\citenamefont {{Iba{\~n}ez-Azpiroz}}, \citenamefont {Tsirkin},\
  and\ \citenamefont {Souza}}]{ibanez-azpiroz_initio_2018}%
  \BibitemOpen
  \bibfield  {author} {\bibinfo {author} {\bibfnamefont {J.}~\bibnamefont
  {{Iba{\~n}ez-Azpiroz}}}, \bibinfo {author} {\bibfnamefont {S.~S.}\
  \bibnamefont {Tsirkin}},\ and\ \bibinfo {author} {\bibfnamefont
  {I.}~\bibnamefont {Souza}},\ }\bibfield  {title} {\bibinfo {title} {{\emph{Ab
  Initio}} calculation of the shift photocurrent by {{Wannier}}
  interpolation},\ }\href {https://doi.org/10.1103/PhysRevB.97.245143}
  {\bibfield  {journal} {\bibinfo  {journal} {Phys. Rev. B}\ }\textbf {\bibinfo
  {volume} {97}},\ \bibinfo {pages} {245143} (\bibinfo {year}
  {2018})}\BibitemShut {NoStop}%
\bibitem [{\citenamefont {Kumbhakar}\ \emph {et~al.}(2021)\citenamefont
  {Kumbhakar}, \citenamefont {Roy~Karmakar}, \citenamefont {Das}, \citenamefont
  {Chakraborty}, \citenamefont {Tiwary},\ and\ \citenamefont
  {Kumbhakar}}]{kumbhakar_reversible_2021}%
  \BibitemOpen
  \bibfield  {author} {\bibinfo {author} {\bibfnamefont {P.}~\bibnamefont
  {Kumbhakar}}, \bibinfo {author} {\bibfnamefont {A.}~\bibnamefont
  {Roy~Karmakar}}, \bibinfo {author} {\bibfnamefont {G.~P.}\ \bibnamefont
  {Das}}, \bibinfo {author} {\bibfnamefont {J.}~\bibnamefont {Chakraborty}},
  \bibinfo {author} {\bibfnamefont {C.~S.}\ \bibnamefont {Tiwary}},\ and\
  \bibinfo {author} {\bibfnamefont {P.}~\bibnamefont {Kumbhakar}},\ }\bibfield
  {title} {\bibinfo {title} {Reversible temperature-dependent photoluminescence
  in semiconductor quantum dots for the development of a smartphone-based
  optical thermometer},\ }\href {https://doi.org/10.1039/D0NR07874C} {\bibfield
   {journal} {\bibinfo  {journal} {Nanoscale}\ }\textbf {\bibinfo {volume}
  {13}},\ \bibinfo {pages} {2946} (\bibinfo {year} {2021})}\BibitemShut
  {NoStop}%
\bibitem [{\citenamefont {Ni}\ \emph {et~al.}(2021)\citenamefont {Ni},
  \citenamefont {Wang}, \citenamefont {Zhang}, \citenamefont {Pozo},
  \citenamefont {Xu}, \citenamefont {Han}, \citenamefont {Manna}, \citenamefont
  {Paglione}, \citenamefont {Felser}, \citenamefont {Grushin}, \citenamefont
  {De~Juan}, \citenamefont {Mele},\ and\ \citenamefont {Wu}}]{ni_giant_2021}%
  \BibitemOpen
  \bibfield  {author} {\bibinfo {author} {\bibfnamefont {Z.}~\bibnamefont
  {Ni}}, \bibinfo {author} {\bibfnamefont {K.}~\bibnamefont {Wang}}, \bibinfo
  {author} {\bibfnamefont {Y.}~\bibnamefont {Zhang}}, \bibinfo {author}
  {\bibfnamefont {O.}~\bibnamefont {Pozo}}, \bibinfo {author} {\bibfnamefont
  {B.}~\bibnamefont {Xu}}, \bibinfo {author} {\bibfnamefont {X.}~\bibnamefont
  {Han}}, \bibinfo {author} {\bibfnamefont {K.}~\bibnamefont {Manna}}, \bibinfo
  {author} {\bibfnamefont {J.}~\bibnamefont {Paglione}}, \bibinfo {author}
  {\bibfnamefont {C.}~\bibnamefont {Felser}}, \bibinfo {author} {\bibfnamefont
  {A.~G.}\ \bibnamefont {Grushin}}, \bibinfo {author} {\bibfnamefont
  {F.}~\bibnamefont {De~Juan}}, \bibinfo {author} {\bibfnamefont {E.~J.}\
  \bibnamefont {Mele}},\ and\ \bibinfo {author} {\bibfnamefont
  {L.}~\bibnamefont {Wu}},\ }\bibfield  {title} {\bibinfo {title} {Giant
  topological longitudinal circular photo-galvanic effect in the chiral
  multifold semimetal {{CoSi}}},\ }\href
  {https://doi.org/10.1038/s41467-020-20408-5} {\bibfield  {journal} {\bibinfo
  {journal} {Nat Commun}\ }\textbf {\bibinfo {volume} {12}},\ \bibinfo {pages}
  {154} (\bibinfo {year} {2021})}\BibitemShut {NoStop}%
\bibitem [{\citenamefont {Ni}\ \emph {et~al.}(2020)\citenamefont {Ni},
  \citenamefont {Xu}, \citenamefont {{S{\'a}nchez-Mart{\'i}nez}}, \citenamefont
  {Zhang}, \citenamefont {Manna}, \citenamefont {Bernhard}, \citenamefont
  {Venderbos}, \citenamefont {De~Juan}, \citenamefont {Felser}, \citenamefont
  {Grushin},\ and\ \citenamefont {Wu}}]{ni_linear_2020}%
  \BibitemOpen
  \bibfield  {author} {\bibinfo {author} {\bibfnamefont {Z.}~\bibnamefont
  {Ni}}, \bibinfo {author} {\bibfnamefont {B.}~\bibnamefont {Xu}}, \bibinfo
  {author} {\bibfnamefont {M.-{\'A}.}\ \bibnamefont
  {{S{\'a}nchez-Mart{\'i}nez}}}, \bibinfo {author} {\bibfnamefont
  {Y.}~\bibnamefont {Zhang}}, \bibinfo {author} {\bibfnamefont
  {K.}~\bibnamefont {Manna}}, \bibinfo {author} {\bibfnamefont
  {C.}~\bibnamefont {Bernhard}}, \bibinfo {author} {\bibfnamefont {J.~W.~F.}\
  \bibnamefont {Venderbos}}, \bibinfo {author} {\bibfnamefont {F.}~\bibnamefont
  {De~Juan}}, \bibinfo {author} {\bibfnamefont {C.}~\bibnamefont {Felser}},
  \bibinfo {author} {\bibfnamefont {A.~G.}\ \bibnamefont {Grushin}},\ and\
  \bibinfo {author} {\bibfnamefont {L.}~\bibnamefont {Wu}},\ }\bibfield
  {title} {\bibinfo {title} {Linear and nonlinear optical responses in the
  chiral multifold semimetal {{RhSi}}},\ }\href
  {https://doi.org/10.1038/s41535-020-00298-y} {\bibfield  {journal} {\bibinfo
  {journal} {npj Quantum Mater.}\ }\textbf {\bibinfo {volume} {5}},\ \bibinfo
  {pages} {96} (\bibinfo {year} {2020})}\BibitemShut {NoStop}%
\bibitem [{\citenamefont {Le}\ \emph {et~al.}(2020)\citenamefont {Le},
  \citenamefont {Zhang}, \citenamefont {Felser},\ and\ \citenamefont
  {Sun}}]{le_initio_2020}%
  \BibitemOpen
  \bibfield  {author} {\bibinfo {author} {\bibfnamefont {C.}~\bibnamefont
  {Le}}, \bibinfo {author} {\bibfnamefont {Y.}~\bibnamefont {Zhang}}, \bibinfo
  {author} {\bibfnamefont {C.}~\bibnamefont {Felser}},\ and\ \bibinfo {author}
  {\bibfnamefont {Y.}~\bibnamefont {Sun}},\ }\bibfield  {title} {\bibinfo
  {title} {{\emph{Ab Initio}} study of quantized circular photogalvanic effect
  in chiral multifold semimetals},\ }\href
  {https://doi.org/10.1103/PhysRevB.102.121111} {\bibfield  {journal} {\bibinfo
   {journal} {Phys. Rev. B}\ }\textbf {\bibinfo {volume} {102}},\ \bibinfo
  {pages} {121111} (\bibinfo {year} {2020})}\BibitemShut {NoStop}%
\bibitem [{\citenamefont {Wang}\ and\ \citenamefont
  {Qian}(2019)}]{wang_ferroicitydriven_2019}%
  \BibitemOpen
  \bibfield  {author} {\bibinfo {author} {\bibfnamefont {H.}~\bibnamefont
  {Wang}}\ and\ \bibinfo {author} {\bibfnamefont {X.}~\bibnamefont {Qian}},\
  }\bibfield  {title} {\bibinfo {title} {Ferroicity-driven nonlinear
  photocurrent switching in time-reversal invariant ferroic materials},\ }\href
  {https://doi.org/10.1126/sciadv.aav9743} {\bibfield  {journal} {\bibinfo
  {journal} {Sci. Adv.}\ }\textbf {\bibinfo {volume} {5}},\ \bibinfo {pages}
  {eaav9743} (\bibinfo {year} {2019})}\BibitemShut {NoStop}%
\bibitem [{\citenamefont {Holder}\ \emph {et~al.}(2020)\citenamefont {Holder},
  \citenamefont {Kaplan},\ and\ \citenamefont
  {Yan}}]{holder_consequences_2020}%
  \BibitemOpen
  \bibfield  {author} {\bibinfo {author} {\bibfnamefont {T.}~\bibnamefont
  {Holder}}, \bibinfo {author} {\bibfnamefont {D.}~\bibnamefont {Kaplan}},\
  and\ \bibinfo {author} {\bibfnamefont {B.}~\bibnamefont {Yan}},\ }\bibfield
  {title} {\bibinfo {title} {Consequences of time-reversal-symmetry breaking in
  the light-matter interaction: {{Berry}} curvature, quantum metric, and
  diabatic motion},\ }\href {https://doi.org/10.1103/PhysRevResearch.2.033100}
  {\bibfield  {journal} {\bibinfo  {journal} {Phys. Rev. Research}\ }\textbf
  {\bibinfo {volume} {2}},\ \bibinfo {pages} {033100} (\bibinfo {year}
  {2020})}\BibitemShut {NoStop}%
\bibitem [{\citenamefont {Sadhukhan}\ and\ \citenamefont
  {Nag}(2021{\natexlab{b}})}]{sadhukhan_role_2021}%
  \BibitemOpen
  \bibfield  {author} {\bibinfo {author} {\bibfnamefont {B.}~\bibnamefont
  {Sadhukhan}}\ and\ \bibinfo {author} {\bibfnamefont {T.}~\bibnamefont
  {Nag}},\ }\bibfield  {title} {\bibinfo {title} {Role of time reversal
  symmetry and tilting in circular photogalvanic responses},\ }\href
  {https://doi.org/10.1103/PhysRevB.103.144308} {\bibfield  {journal} {\bibinfo
   {journal} {Phys. Rev. B}\ }\textbf {\bibinfo {volume} {103}},\ \bibinfo
  {pages} {144308} (\bibinfo {year} {2021}{\natexlab{b}})}\BibitemShut
  {NoStop}%
\bibitem [{\citenamefont {Yang}\ \emph {et~al.}(2021)\citenamefont {Yang},
  \citenamefont {Singh}, \citenamefont {Gaudet}, \citenamefont {Lu},
  \citenamefont {Huang}, \citenamefont {Chiu}, \citenamefont {Huang},
  \citenamefont {Wang}, \citenamefont {Bahrami}, \citenamefont {Xu},
  \citenamefont {Franklin}, \citenamefont {Sochnikov}, \citenamefont {Graf},
  \citenamefont {Xu}, \citenamefont {Zhao}, \citenamefont {Hoffman},
  \citenamefont {Lin}, \citenamefont {Torchinsky}, \citenamefont {Broholm},
  \citenamefont {Bansil},\ and\ \citenamefont
  {Tafti}}]{yang_noncollinear_2021}%
  \BibitemOpen
  \bibfield  {author} {\bibinfo {author} {\bibfnamefont {H.-Y.}\ \bibnamefont
  {Yang}}, \bibinfo {author} {\bibfnamefont {B.}~\bibnamefont {Singh}},
  \bibinfo {author} {\bibfnamefont {J.}~\bibnamefont {Gaudet}}, \bibinfo
  {author} {\bibfnamefont {B.}~\bibnamefont {Lu}}, \bibinfo {author}
  {\bibfnamefont {C.-Y.}\ \bibnamefont {Huang}}, \bibinfo {author}
  {\bibfnamefont {W.-C.}\ \bibnamefont {Chiu}}, \bibinfo {author}
  {\bibfnamefont {S.-M.}\ \bibnamefont {Huang}}, \bibinfo {author}
  {\bibfnamefont {B.}~\bibnamefont {Wang}}, \bibinfo {author} {\bibfnamefont
  {F.}~\bibnamefont {Bahrami}}, \bibinfo {author} {\bibfnamefont
  {B.}~\bibnamefont {Xu}}, \bibinfo {author} {\bibfnamefont {J.}~\bibnamefont
  {Franklin}}, \bibinfo {author} {\bibfnamefont {I.}~\bibnamefont {Sochnikov}},
  \bibinfo {author} {\bibfnamefont {D.~E.}\ \bibnamefont {Graf}}, \bibinfo
  {author} {\bibfnamefont {G.}~\bibnamefont {Xu}}, \bibinfo {author}
  {\bibfnamefont {Y.}~\bibnamefont {Zhao}}, \bibinfo {author} {\bibfnamefont
  {C.~M.}\ \bibnamefont {Hoffman}}, \bibinfo {author} {\bibfnamefont
  {H.}~\bibnamefont {Lin}}, \bibinfo {author} {\bibfnamefont {D.~H.}\
  \bibnamefont {Torchinsky}}, \bibinfo {author} {\bibfnamefont {C.~L.}\
  \bibnamefont {Broholm}}, \bibinfo {author} {\bibfnamefont {A.}~\bibnamefont
  {Bansil}},\ and\ \bibinfo {author} {\bibfnamefont {F.}~\bibnamefont
  {Tafti}},\ }\bibfield  {title} {\bibinfo {title} {Noncollinear ferromagnetic
  {{Weyl}} semimetal with anisotropic anomalous {{Hall}} effect},\ }\href
  {https://doi.org/10.1103/PhysRevB.103.115143} {\bibfield  {journal} {\bibinfo
   {journal} {Phys. Rev. B}\ }\textbf {\bibinfo {volume} {103}},\ \bibinfo
  {pages} {115143} (\bibinfo {year} {2021})}\BibitemShut {NoStop}%
\bibitem [{\citenamefont {Alam}\ \emph {et~al.}(2023)\citenamefont {Alam},
  \citenamefont {Fakhredine}, \citenamefont {Ahmad}, \citenamefont {Tanwar},
  \citenamefont {Yang}, \citenamefont {Tafti}, \citenamefont {Cuono},
  \citenamefont {Islam}, \citenamefont {Singh}, \citenamefont {Lynnyk},
  \citenamefont {Autieri},\ and\ \citenamefont {Matusiak}}]{alam_sign_2023}%
  \BibitemOpen
  \bibfield  {author} {\bibinfo {author} {\bibfnamefont {M.~S.}\ \bibnamefont
  {Alam}}, \bibinfo {author} {\bibfnamefont {A.}~\bibnamefont {Fakhredine}},
  \bibinfo {author} {\bibfnamefont {M.}~\bibnamefont {Ahmad}}, \bibinfo
  {author} {\bibfnamefont {P.~K.}\ \bibnamefont {Tanwar}}, \bibinfo {author}
  {\bibfnamefont {H.-Y.}\ \bibnamefont {Yang}}, \bibinfo {author}
  {\bibfnamefont {F.}~\bibnamefont {Tafti}}, \bibinfo {author} {\bibfnamefont
  {G.}~\bibnamefont {Cuono}}, \bibinfo {author} {\bibfnamefont
  {R.}~\bibnamefont {Islam}}, \bibinfo {author} {\bibfnamefont
  {B.}~\bibnamefont {Singh}}, \bibinfo {author} {\bibfnamefont
  {A.}~\bibnamefont {Lynnyk}}, \bibinfo {author} {\bibfnamefont
  {C.}~\bibnamefont {Autieri}},\ and\ \bibinfo {author} {\bibfnamefont
  {M.}~\bibnamefont {Matusiak}},\ }\bibfield  {title} {\bibinfo {title} {Sign
  change of anomalous {{Hall}} effect and anomalous {{Nernst}} effect in the
  {{Weyl}} semimetal {{CeAlSi}}},\ }\href
  {https://doi.org/10.1103/PhysRevB.107.085102} {\bibfield  {journal} {\bibinfo
   {journal} {Phys. Rev. B}\ }\textbf {\bibinfo {volume} {107}},\ \bibinfo
  {pages} {085102} (\bibinfo {year} {2023})}\BibitemShut {NoStop}%
\bibitem [{\citenamefont {Piva}\ \emph {et~al.}(2023)\citenamefont {Piva},
  \citenamefont {Souza}, \citenamefont {{Brousseau-Couture}}, \citenamefont
  {Sorn}, \citenamefont {Pakuszewski}, \citenamefont {John}, \citenamefont
  {Adriano}, \citenamefont {C{\^o}t{\'e}}, \citenamefont {Pagliuso},
  \citenamefont {Paramekanti},\ and\ \citenamefont
  {Nicklas}}]{piva_topological_2023}%
  \BibitemOpen
  \bibfield  {author} {\bibinfo {author} {\bibfnamefont {M.~M.}\ \bibnamefont
  {Piva}}, \bibinfo {author} {\bibfnamefont {J.~C.}\ \bibnamefont {Souza}},
  \bibinfo {author} {\bibfnamefont {V.}~\bibnamefont {{Brousseau-Couture}}},
  \bibinfo {author} {\bibfnamefont {S.}~\bibnamefont {Sorn}}, \bibinfo {author}
  {\bibfnamefont {K.~R.}\ \bibnamefont {Pakuszewski}}, \bibinfo {author}
  {\bibfnamefont {J.~K.}\ \bibnamefont {John}}, \bibinfo {author}
  {\bibfnamefont {C.}~\bibnamefont {Adriano}}, \bibinfo {author} {\bibfnamefont
  {M.}~\bibnamefont {C{\^o}t{\'e}}}, \bibinfo {author} {\bibfnamefont {P.~G.}\
  \bibnamefont {Pagliuso}}, \bibinfo {author} {\bibfnamefont {A.}~\bibnamefont
  {Paramekanti}},\ and\ \bibinfo {author} {\bibfnamefont {M.}~\bibnamefont
  {Nicklas}},\ }\bibfield  {title} {\bibinfo {title} {Topological features in
  the ferromagnetic {{Weyl}} semimetal {{CeAlSi}}: {{Role}} of domain walls},\
  }\href {https://doi.org/10.1103/PhysRevResearch.5.013068} {\bibfield
  {journal} {\bibinfo  {journal} {Phys. Rev. Research}\ }\textbf {\bibinfo
  {volume} {5}},\ \bibinfo {pages} {013068} (\bibinfo {year}
  {2023})}\BibitemShut {NoStop}%
\bibitem [{\citenamefont {Sakhya}\ \emph {et~al.}(2023)\citenamefont {Sakhya},
  \citenamefont {Huang}, \citenamefont {Dhakal}, \citenamefont {Gao},
  \citenamefont {Regmi}, \citenamefont {Wang}, \citenamefont {Wen},
  \citenamefont {He}, \citenamefont {Yao}, \citenamefont {Smith}, \citenamefont
  {Sprague}, \citenamefont {Gao}, \citenamefont {Singh}, \citenamefont {Lin},
  \citenamefont {Xu}, \citenamefont {Tafti}, \citenamefont {Bansil},\ and\
  \citenamefont {Neupane}}]{sakhya_observation_2023}%
  \BibitemOpen
  \bibfield  {author} {\bibinfo {author} {\bibfnamefont {A.~P.}\ \bibnamefont
  {Sakhya}}, \bibinfo {author} {\bibfnamefont {C.-Y.}\ \bibnamefont {Huang}},
  \bibinfo {author} {\bibfnamefont {G.}~\bibnamefont {Dhakal}}, \bibinfo
  {author} {\bibfnamefont {X.-J.}\ \bibnamefont {Gao}}, \bibinfo {author}
  {\bibfnamefont {S.}~\bibnamefont {Regmi}}, \bibinfo {author} {\bibfnamefont
  {B.}~\bibnamefont {Wang}}, \bibinfo {author} {\bibfnamefont {W.}~\bibnamefont
  {Wen}}, \bibinfo {author} {\bibfnamefont {R.-H.}\ \bibnamefont {He}},
  \bibinfo {author} {\bibfnamefont {X.}~\bibnamefont {Yao}}, \bibinfo {author}
  {\bibfnamefont {R.}~\bibnamefont {Smith}}, \bibinfo {author} {\bibfnamefont
  {M.}~\bibnamefont {Sprague}}, \bibinfo {author} {\bibfnamefont
  {S.}~\bibnamefont {Gao}}, \bibinfo {author} {\bibfnamefont {B.}~\bibnamefont
  {Singh}}, \bibinfo {author} {\bibfnamefont {H.}~\bibnamefont {Lin}}, \bibinfo
  {author} {\bibfnamefont {S.-Y.}\ \bibnamefont {Xu}}, \bibinfo {author}
  {\bibfnamefont {F.}~\bibnamefont {Tafti}}, \bibinfo {author} {\bibfnamefont
  {A.}~\bibnamefont {Bansil}},\ and\ \bibinfo {author} {\bibfnamefont
  {M.}~\bibnamefont {Neupane}},\ }\bibfield  {title} {\bibinfo {title}
  {Observation of {{Fermi}} arcs and {{Weyl}} nodes in a noncentrosymmetric
  magnetic {{Weyl}} semimetal},\ }\href
  {https://doi.org/10.1103/PhysRevMaterials.7.L051202} {\bibfield  {journal}
  {\bibinfo  {journal} {Phys. Rev. Materials}\ }\textbf {\bibinfo {volume}
  {7}},\ \bibinfo {pages} {L051202} (\bibinfo {year} {2023})}\BibitemShut
  {NoStop}%
\bibitem [{\citenamefont {Zhang}\ \emph {et~al.}(2023)\citenamefont {Zhang},
  \citenamefont {Ding}, \citenamefont {Zhan}, \citenamefont {Li}, \citenamefont
  {Li}, \citenamefont {Tang}, \citenamefont {Qian}, \citenamefont {Pan},
  \citenamefont {Xiao}, \citenamefont {Zhang}, \citenamefont {Wang},
  \citenamefont {Xiang},\ and\ \citenamefont
  {Chen}}]{zhang_temperaturedependent_2023}%
  \BibitemOpen
  \bibfield  {author} {\bibinfo {author} {\bibfnamefont {N.}~\bibnamefont
  {Zhang}}, \bibinfo {author} {\bibfnamefont {X.}~\bibnamefont {Ding}},
  \bibinfo {author} {\bibfnamefont {F.}~\bibnamefont {Zhan}}, \bibinfo {author}
  {\bibfnamefont {H.}~\bibnamefont {Li}}, \bibinfo {author} {\bibfnamefont
  {H.}~\bibnamefont {Li}}, \bibinfo {author} {\bibfnamefont {K.}~\bibnamefont
  {Tang}}, \bibinfo {author} {\bibfnamefont {Y.}~\bibnamefont {Qian}}, \bibinfo
  {author} {\bibfnamefont {S.}~\bibnamefont {Pan}}, \bibinfo {author}
  {\bibfnamefont {X.}~\bibnamefont {Xiao}}, \bibinfo {author} {\bibfnamefont
  {J.}~\bibnamefont {Zhang}}, \bibinfo {author} {\bibfnamefont
  {R.}~\bibnamefont {Wang}}, \bibinfo {author} {\bibfnamefont {Z.}~\bibnamefont
  {Xiang}},\ and\ \bibinfo {author} {\bibfnamefont {X.}~\bibnamefont {Chen}},\
  }\bibfield  {title} {\bibinfo {title} {Temperature-dependent and
  magnetism-controlled {{Fermi}} surface changes in magnetic {{Weyl}}
  semimetals},\ }\href {https://doi.org/10.1103/PhysRevResearch.5.L022013}
  {\bibfield  {journal} {\bibinfo  {journal} {Phys. Rev. Research}\ }\textbf
  {\bibinfo {volume} {5}},\ \bibinfo {pages} {L022013} (\bibinfo {year}
  {2023})}\BibitemShut {NoStop}%
\bibitem [{\citenamefont {Chang}\ \emph
  {et~al.}(2018{\natexlab{b}})\citenamefont {Chang}, \citenamefont {Singh},
  \citenamefont {Xu}, \citenamefont {Bian}, \citenamefont {Huang},
  \citenamefont {Hsu}, \citenamefont {Belopolski}, \citenamefont {Alidoust},
  \citenamefont {Sanchez}, \citenamefont {Zheng}, \citenamefont {Lu},
  \citenamefont {Zhang}, \citenamefont {Bian}, \citenamefont {Chang},
  \citenamefont {Jeng}, \citenamefont {Bansil}, \citenamefont {Hsu},
  \citenamefont {Jia}, \citenamefont {Neupert}, \citenamefont {Lin},\ and\
  \citenamefont {Hasan}}]{chang_magnetic_2018}%
  \BibitemOpen
  \bibfield  {author} {\bibinfo {author} {\bibfnamefont {G.}~\bibnamefont
  {Chang}}, \bibinfo {author} {\bibfnamefont {B.}~\bibnamefont {Singh}},
  \bibinfo {author} {\bibfnamefont {S.-Y.}\ \bibnamefont {Xu}}, \bibinfo
  {author} {\bibfnamefont {G.}~\bibnamefont {Bian}}, \bibinfo {author}
  {\bibfnamefont {S.-M.}\ \bibnamefont {Huang}}, \bibinfo {author}
  {\bibfnamefont {C.-H.}\ \bibnamefont {Hsu}}, \bibinfo {author} {\bibfnamefont
  {I.}~\bibnamefont {Belopolski}}, \bibinfo {author} {\bibfnamefont
  {N.}~\bibnamefont {Alidoust}}, \bibinfo {author} {\bibfnamefont {D.~S.}\
  \bibnamefont {Sanchez}}, \bibinfo {author} {\bibfnamefont {H.}~\bibnamefont
  {Zheng}}, \bibinfo {author} {\bibfnamefont {H.}~\bibnamefont {Lu}}, \bibinfo
  {author} {\bibfnamefont {X.}~\bibnamefont {Zhang}}, \bibinfo {author}
  {\bibfnamefont {Y.}~\bibnamefont {Bian}}, \bibinfo {author} {\bibfnamefont
  {T.-R.}\ \bibnamefont {Chang}}, \bibinfo {author} {\bibfnamefont {H.-T.}\
  \bibnamefont {Jeng}}, \bibinfo {author} {\bibfnamefont {A.}~\bibnamefont
  {Bansil}}, \bibinfo {author} {\bibfnamefont {H.}~\bibnamefont {Hsu}},
  \bibinfo {author} {\bibfnamefont {S.}~\bibnamefont {Jia}}, \bibinfo {author}
  {\bibfnamefont {T.}~\bibnamefont {Neupert}}, \bibinfo {author} {\bibfnamefont
  {H.}~\bibnamefont {Lin}},\ and\ \bibinfo {author} {\bibfnamefont {M.~Z.}\
  \bibnamefont {Hasan}},\ }\bibfield  {title} {\bibinfo {title} {Magnetic and
  noncentrosymmetric {{Weyl}} fermion semimetals in the {{R AlGe}} family of
  compounds ( {{R}} = rare earth )},\ }\href
  {https://doi.org/10.1103/PhysRevB.97.041104} {\bibfield  {journal} {\bibinfo
  {journal} {Phys. Rev. B}\ }\textbf {\bibinfo {volume} {97}},\ \bibinfo
  {pages} {041104} (\bibinfo {year} {2018}{\natexlab{b}})}\BibitemShut
  {NoStop}%
\bibitem [{\citenamefont {Xu}\ \emph {et~al.}(2017)\citenamefont {Xu},
  \citenamefont {Alidoust}, \citenamefont {Chang}, \citenamefont {Lu},
  \citenamefont {Singh}, \citenamefont {Belopolski}, \citenamefont {Sanchez},
  \citenamefont {Zhang}, \citenamefont {Bian}, \citenamefont {Zheng},
  \citenamefont {Husanu}, \citenamefont {Bian}, \citenamefont {Huang},
  \citenamefont {Hsu}, \citenamefont {Chang}, \citenamefont {Jeng},
  \citenamefont {Bansil}, \citenamefont {Neupert}, \citenamefont {Strocov},
  \citenamefont {Lin}, \citenamefont {Jia},\ and\ \citenamefont
  {Hasan}}]{xu_discovery_2017}%
  \BibitemOpen
  \bibfield  {author} {\bibinfo {author} {\bibfnamefont {S.-Y.}\ \bibnamefont
  {Xu}}, \bibinfo {author} {\bibfnamefont {N.}~\bibnamefont {Alidoust}},
  \bibinfo {author} {\bibfnamefont {G.}~\bibnamefont {Chang}}, \bibinfo
  {author} {\bibfnamefont {H.}~\bibnamefont {Lu}}, \bibinfo {author}
  {\bibfnamefont {B.}~\bibnamefont {Singh}}, \bibinfo {author} {\bibfnamefont
  {I.}~\bibnamefont {Belopolski}}, \bibinfo {author} {\bibfnamefont {D.~S.}\
  \bibnamefont {Sanchez}}, \bibinfo {author} {\bibfnamefont {X.}~\bibnamefont
  {Zhang}}, \bibinfo {author} {\bibfnamefont {G.}~\bibnamefont {Bian}},
  \bibinfo {author} {\bibfnamefont {H.}~\bibnamefont {Zheng}}, \bibinfo
  {author} {\bibfnamefont {M.-A.}\ \bibnamefont {Husanu}}, \bibinfo {author}
  {\bibfnamefont {Y.}~\bibnamefont {Bian}}, \bibinfo {author} {\bibfnamefont
  {S.-M.}\ \bibnamefont {Huang}}, \bibinfo {author} {\bibfnamefont {C.-H.}\
  \bibnamefont {Hsu}}, \bibinfo {author} {\bibfnamefont {T.-R.}\ \bibnamefont
  {Chang}}, \bibinfo {author} {\bibfnamefont {H.-T.}\ \bibnamefont {Jeng}},
  \bibinfo {author} {\bibfnamefont {A.}~\bibnamefont {Bansil}}, \bibinfo
  {author} {\bibfnamefont {T.}~\bibnamefont {Neupert}}, \bibinfo {author}
  {\bibfnamefont {V.~N.}\ \bibnamefont {Strocov}}, \bibinfo {author}
  {\bibfnamefont {H.}~\bibnamefont {Lin}}, \bibinfo {author} {\bibfnamefont
  {S.}~\bibnamefont {Jia}},\ and\ \bibinfo {author} {\bibfnamefont {M.~Z.}\
  \bibnamefont {Hasan}},\ }\bibfield  {title} {\bibinfo {title} {Discovery of
  {{Lorentz-violating}} type {{II Weyl}} fermions in {{LaAlGe}}},\ }\href
  {https://doi.org/10.1126/sciadv.1603266} {\bibfield  {journal} {\bibinfo
  {journal} {Sci. Adv.}\ }\textbf {\bibinfo {volume} {3}},\ \bibinfo {pages}
  {e1603266} (\bibinfo {year} {2017})}\BibitemShut {NoStop}%
\bibitem [{\citenamefont {Kresse}\ and\ \citenamefont
  {Furthm{\"u}ller}(1996)}]{kresse_efficient_1996}%
  \BibitemOpen
  \bibfield  {author} {\bibinfo {author} {\bibfnamefont {G.}~\bibnamefont
  {Kresse}}\ and\ \bibinfo {author} {\bibfnamefont {J.}~\bibnamefont
  {Furthm{\"u}ller}},\ }\bibfield  {title} {\bibinfo {title} {Efficient
  iterative schemes for {\emph{ab initio}} total-energy calculations using a
  plane-wave basis set},\ }\href {https://doi.org/10.1103/PhysRevB.54.11169}
  {\bibfield  {journal} {\bibinfo  {journal} {Phys. Rev. B}\ }\textbf {\bibinfo
  {volume} {54}},\ \bibinfo {pages} {11169} (\bibinfo {year}
  {1996})}\BibitemShut {NoStop}%
\bibitem [{\citenamefont {Perdew}\ \emph {et~al.}(1996)\citenamefont {Perdew},
  \citenamefont {Burke},\ and\ \citenamefont
  {Ernzerhof}}]{perdew_generalized_1996}%
  \BibitemOpen
  \bibfield  {author} {\bibinfo {author} {\bibfnamefont {J.~P.}\ \bibnamefont
  {Perdew}}, \bibinfo {author} {\bibfnamefont {K.}~\bibnamefont {Burke}},\ and\
  \bibinfo {author} {\bibfnamefont {M.}~\bibnamefont {Ernzerhof}},\ }\bibfield
  {title} {\bibinfo {title} {Generalized {{Gradient Approximation Made
  Simple}}},\ }\href {https://doi.org/10.1103/PhysRevLett.77.3865} {\bibfield
  {journal} {\bibinfo  {journal} {Phys. Rev. Lett.}\ }\textbf {\bibinfo
  {volume} {77}},\ \bibinfo {pages} {3865} (\bibinfo {year}
  {1996})}\BibitemShut {NoStop}%
\bibitem [{\citenamefont {Dudarev}\ \emph {et~al.}(1998)\citenamefont
  {Dudarev}, \citenamefont {Botton}, \citenamefont {Savrasov}, \citenamefont
  {Humphreys},\ and\ \citenamefont {Sutton}}]{dudarev_electronenergyloss_1998}%
  \BibitemOpen
  \bibfield  {author} {\bibinfo {author} {\bibfnamefont {S.~L.}\ \bibnamefont
  {Dudarev}}, \bibinfo {author} {\bibfnamefont {G.~A.}\ \bibnamefont {Botton}},
  \bibinfo {author} {\bibfnamefont {S.~Y.}\ \bibnamefont {Savrasov}}, \bibinfo
  {author} {\bibfnamefont {C.~J.}\ \bibnamefont {Humphreys}},\ and\ \bibinfo
  {author} {\bibfnamefont {A.~P.}\ \bibnamefont {Sutton}},\ }\bibfield  {title}
  {\bibinfo {title} {Electron-energy-loss spectra and the structural stability
  of nickel oxide: {{An LSDA}}+{{U}} study},\ }\href
  {https://doi.org/10.1103/PhysRevB.57.1505} {\bibfield  {journal} {\bibinfo
  {journal} {Phys. Rev. B}\ }\textbf {\bibinfo {volume} {57}},\ \bibinfo
  {pages} {1505} (\bibinfo {year} {1998})}\BibitemShut {NoStop}%
\bibitem [{\citenamefont {Pizzi}\ \emph {et~al.}(2020)\citenamefont {Pizzi},
  \citenamefont {Vitale}, \citenamefont {Arita}, \citenamefont {Bl{\"u}gel},
  \citenamefont {Freimuth}, \citenamefont {G{\'e}ranton}, \citenamefont
  {Gibertini}, \citenamefont {Gresch}, \citenamefont {Johnson}, \citenamefont
  {Koretsune}, \citenamefont {{Iba{\~n}ez-Azpiroz}}, \citenamefont {Lee},
  \citenamefont {Lihm}, \citenamefont {Marchand}, \citenamefont {Marrazzo},
  \citenamefont {Mokrousov}, \citenamefont {Mustafa}, \citenamefont {Nohara},
  \citenamefont {Nomura}, \citenamefont {Paulatto}, \citenamefont {Ponc{\'e}},
  \citenamefont {Ponweiser}, \citenamefont {Qiao}, \citenamefont {Th{\"o}le},
  \citenamefont {Tsirkin}, \citenamefont {Wierzbowska}, \citenamefont
  {Marzari}, \citenamefont {Vanderbilt}, \citenamefont {Souza}, \citenamefont
  {Mostofi},\ and\ \citenamefont {Yates}}]{pizzi_wannier90_2020}%
  \BibitemOpen
  \bibfield  {author} {\bibinfo {author} {\bibfnamefont {G.}~\bibnamefont
  {Pizzi}}, \bibinfo {author} {\bibfnamefont {V.}~\bibnamefont {Vitale}},
  \bibinfo {author} {\bibfnamefont {R.}~\bibnamefont {Arita}}, \bibinfo
  {author} {\bibfnamefont {S.}~\bibnamefont {Bl{\"u}gel}}, \bibinfo {author}
  {\bibfnamefont {F.}~\bibnamefont {Freimuth}}, \bibinfo {author}
  {\bibfnamefont {G.}~\bibnamefont {G{\'e}ranton}}, \bibinfo {author}
  {\bibfnamefont {M.}~\bibnamefont {Gibertini}}, \bibinfo {author}
  {\bibfnamefont {D.}~\bibnamefont {Gresch}}, \bibinfo {author} {\bibfnamefont
  {C.}~\bibnamefont {Johnson}}, \bibinfo {author} {\bibfnamefont
  {T.}~\bibnamefont {Koretsune}}, \bibinfo {author} {\bibfnamefont
  {J.}~\bibnamefont {{Iba{\~n}ez-Azpiroz}}}, \bibinfo {author} {\bibfnamefont
  {H.}~\bibnamefont {Lee}}, \bibinfo {author} {\bibfnamefont {J.-M.}\
  \bibnamefont {Lihm}}, \bibinfo {author} {\bibfnamefont {D.}~\bibnamefont
  {Marchand}}, \bibinfo {author} {\bibfnamefont {A.}~\bibnamefont {Marrazzo}},
  \bibinfo {author} {\bibfnamefont {Y.}~\bibnamefont {Mokrousov}}, \bibinfo
  {author} {\bibfnamefont {J.~I.}\ \bibnamefont {Mustafa}}, \bibinfo {author}
  {\bibfnamefont {Y.}~\bibnamefont {Nohara}}, \bibinfo {author} {\bibfnamefont
  {Y.}~\bibnamefont {Nomura}}, \bibinfo {author} {\bibfnamefont
  {L.}~\bibnamefont {Paulatto}}, \bibinfo {author} {\bibfnamefont
  {S.}~\bibnamefont {Ponc{\'e}}}, \bibinfo {author} {\bibfnamefont
  {T.}~\bibnamefont {Ponweiser}}, \bibinfo {author} {\bibfnamefont
  {J.}~\bibnamefont {Qiao}}, \bibinfo {author} {\bibfnamefont {F.}~\bibnamefont
  {Th{\"o}le}}, \bibinfo {author} {\bibfnamefont {S.~S.}\ \bibnamefont
  {Tsirkin}}, \bibinfo {author} {\bibfnamefont {M.}~\bibnamefont
  {Wierzbowska}}, \bibinfo {author} {\bibfnamefont {N.}~\bibnamefont
  {Marzari}}, \bibinfo {author} {\bibfnamefont {D.}~\bibnamefont {Vanderbilt}},
  \bibinfo {author} {\bibfnamefont {I.}~\bibnamefont {Souza}}, \bibinfo
  {author} {\bibfnamefont {A.~A.}\ \bibnamefont {Mostofi}},\ and\ \bibinfo
  {author} {\bibfnamefont {J.~R.}\ \bibnamefont {Yates}},\ }\bibfield  {title}
  {\bibinfo {title} {Wannier90 as a community code: New features and
  applications},\ }\href {https://doi.org/10.1088/1361-648X/ab51ff} {\bibfield
  {journal} {\bibinfo  {journal} {J. Phys.: Condens. Matter}\ }\textbf
  {\bibinfo {volume} {32}},\ \bibinfo {pages} {165902} (\bibinfo {year}
  {2020})}\BibitemShut {NoStop}%
\bibitem [{\citenamefont {Gresch}\ \emph {et~al.}(2018)\citenamefont {Gresch},
  \citenamefont {Wu}, \citenamefont {Winkler}, \citenamefont {H{\"a}uselmann},
  \citenamefont {Troyer},\ and\ \citenamefont
  {Soluyanov}}]{gresch_automated_2018}%
  \BibitemOpen
  \bibfield  {author} {\bibinfo {author} {\bibfnamefont {D.}~\bibnamefont
  {Gresch}}, \bibinfo {author} {\bibfnamefont {Q.}~\bibnamefont {Wu}}, \bibinfo
  {author} {\bibfnamefont {G.~W.}\ \bibnamefont {Winkler}}, \bibinfo {author}
  {\bibfnamefont {R.}~\bibnamefont {H{\"a}uselmann}}, \bibinfo {author}
  {\bibfnamefont {M.}~\bibnamefont {Troyer}},\ and\ \bibinfo {author}
  {\bibfnamefont {A.~A.}\ \bibnamefont {Soluyanov}},\ }\bibfield  {title}
  {\bibinfo {title} {Automated construction of symmetrized {{Wannier-like}}
  tight-binding models from {\emph{ab initio}} calculations},\ }\href
  {https://doi.org/10.1103/PhysRevMaterials.2.103805} {\bibfield  {journal}
  {\bibinfo  {journal} {Phys. Rev. Materials}\ }\textbf {\bibinfo {volume}
  {2}},\ \bibinfo {pages} {103805} (\bibinfo {year} {2018})}\BibitemShut
  {NoStop}%
\bibitem [{\citenamefont {Tsirkin}(2021)}]{tsirkin_high_2021}%
  \BibitemOpen
  \bibfield  {author} {\bibinfo {author} {\bibfnamefont {S.~S.}\ \bibnamefont
  {Tsirkin}},\ }\bibfield  {title} {\bibinfo {title} {High performance
  {{Wannier}} interpolation of {{Berry}} curvature and related quantities with
  {{WannierBerri}} code},\ }\href {https://doi.org/10.1038/s41524-021-00498-5}
  {\bibfield  {journal} {\bibinfo  {journal} {npj Comput Mater}\ }\textbf
  {\bibinfo {volume} {7}},\ \bibinfo {pages} {33} (\bibinfo {year}
  {2021})}\BibitemShut {NoStop}%
\bibitem [{\citenamefont {Sipe}\ and\ \citenamefont
  {Shkrebtii}(2000)}]{sipe_secondorder_2000}%
  \BibitemOpen
  \bibfield  {author} {\bibinfo {author} {\bibfnamefont {J.~E.}\ \bibnamefont
  {Sipe}}\ and\ \bibinfo {author} {\bibfnamefont {A.~I.}\ \bibnamefont
  {Shkrebtii}},\ }\bibfield  {title} {\bibinfo {title} {Second-order optical
  response in semiconductors},\ }\href
  {https://doi.org/10.1103/PhysRevB.61.5337} {\bibfield  {journal} {\bibinfo
  {journal} {Phys. Rev. B}\ }\textbf {\bibinfo {volume} {61}},\ \bibinfo
  {pages} {5337} (\bibinfo {year} {2000})}\BibitemShut {NoStop}%
\bibitem [{\citenamefont {Zhang}\ \emph {et~al.}(2019)\citenamefont {Zhang},
  \citenamefont {Holder}, \citenamefont {Ishizuka}, \citenamefont {De~Juan},
  \citenamefont {Nagaosa}, \citenamefont {Felser},\ and\ \citenamefont
  {Yan}}]{zhang_switchable_2019}%
  \BibitemOpen
  \bibfield  {author} {\bibinfo {author} {\bibfnamefont {Y.}~\bibnamefont
  {Zhang}}, \bibinfo {author} {\bibfnamefont {T.}~\bibnamefont {Holder}},
  \bibinfo {author} {\bibfnamefont {H.}~\bibnamefont {Ishizuka}}, \bibinfo
  {author} {\bibfnamefont {F.}~\bibnamefont {De~Juan}}, \bibinfo {author}
  {\bibfnamefont {N.}~\bibnamefont {Nagaosa}}, \bibinfo {author} {\bibfnamefont
  {C.}~\bibnamefont {Felser}},\ and\ \bibinfo {author} {\bibfnamefont
  {B.}~\bibnamefont {Yan}},\ }\bibfield  {title} {\bibinfo {title} {Switchable
  magnetic bulk photovoltaic effect in the two-dimensional magnet {{CrI3}}},\
  }\href {https://doi.org/10.1038/s41467-019-11832-3} {\bibfield  {journal}
  {\bibinfo  {journal} {Nat Commun}\ }\textbf {\bibinfo {volume} {10}},\
  \bibinfo {pages} {3783} (\bibinfo {year} {2019})}\BibitemShut {NoStop}%
\bibitem [{\citenamefont {Drueke}(2021)}]{drueke_nonlinear_2021}%
  \BibitemOpen
  \bibfield  {author} {\bibinfo {author} {\bibfnamefont {E.~A.}\ \bibnamefont
  {Drueke}},\ }\emph {\bibinfo {title} {Nonlinear {{Optical Effects}} in {{Weyl
  Semimetals}} and {{Other Strongly Correlated Materials}}}},\ \href@noop {}
  {Ph.D. thesis},\ \bibinfo  {school} {University of Michigan}, \bibinfo
  {address} {USA} (\bibinfo {year} {2021})\BibitemShut {NoStop}%
\bibitem [{\citenamefont {Kaplan}\ \emph {et~al.}(2020)\citenamefont {Kaplan},
  \citenamefont {Holder},\ and\ \citenamefont
  {Yan}}]{kaplan_nonvanishing_2020}%
  \BibitemOpen
  \bibfield  {author} {\bibinfo {author} {\bibfnamefont {D.}~\bibnamefont
  {Kaplan}}, \bibinfo {author} {\bibfnamefont {T.}~\bibnamefont {Holder}},\
  and\ \bibinfo {author} {\bibfnamefont {B.}~\bibnamefont {Yan}},\ }\bibfield
  {title} {\bibinfo {title} {Nonvanishing {{Subgap Photocurrent}} as a
  {{Probe}} of {{Lifetime Effects}}},\ }\href
  {https://doi.org/10.1103/PhysRevLett.125.227401} {\bibfield  {journal}
  {\bibinfo  {journal} {Phys. Rev. Lett.}\ }\textbf {\bibinfo {volume} {125}},\
  \bibinfo {pages} {227401} (\bibinfo {year} {2020})}\BibitemShut {NoStop}%
\bibitem [{\citenamefont {Kaplan}\ \emph {et~al.}(2023)\citenamefont {Kaplan},
  \citenamefont {Holder},\ and\ \citenamefont {Yan}}]{kaplan_unifying_2023}%
  \BibitemOpen
  \bibfield  {author} {\bibinfo {author} {\bibfnamefont {D.}~\bibnamefont
  {Kaplan}}, \bibinfo {author} {\bibfnamefont {T.}~\bibnamefont {Holder}},\
  and\ \bibinfo {author} {\bibfnamefont {B.}~\bibnamefont {Yan}},\ }\bibfield
  {title} {\bibinfo {title} {Unifying semiclassics and quantum perturbation
  theory at nonlinear order},\ }\href
  {https://doi.org/10.21468/SciPostPhys.14.4.082} {\bibfield  {journal}
  {\bibinfo  {journal} {SciPost Phys.}\ }\textbf {\bibinfo {volume} {14}},\
  \bibinfo {pages} {082} (\bibinfo {year} {2023})}\BibitemShut {NoStop}%
\bibitem [{\citenamefont {Butcher}(1965)}]{butcher_nonlinear_1965}%
  \BibitemOpen
  \bibfield  {author} {\bibinfo {author} {\bibfnamefont {P.~N.}\ \bibnamefont
  {Butcher}},\ }\href@noop {} {\emph {\bibinfo {title} {Nonlinear {{Optical
  Phenomena}}}}},\ Vol.\ \bibinfo {volume} {200}\ (\bibinfo  {publisher}
  {Engineering Experiment Station of the Ohio State University},\ \bibinfo
  {year} {1965})\BibitemShut {NoStop}%
\bibitem [{\citenamefont {Boyd}(2020)}]{boyd_nonlinear_2020}%
  \BibitemOpen
  \bibfield  {author} {\bibinfo {author} {\bibfnamefont {R.~W.}\ \bibnamefont
  {Boyd}},\ }\href@noop {} {\emph {\bibinfo {title} {Nonlinear Optics}}},\
  \bibinfo {edition} {fourth edition}\ ed.\ (\bibinfo  {publisher} {Academic
  Press, an imprint of Elsevier},\ \bibinfo {address} {London, United Kingdom ;
  San Diego, CA},\ \bibinfo {year} {2020})\BibitemShut {NoStop}%
\bibitem [{\citenamefont {Litvin}(2013)}]{litvin_magnetic_2013}%
  \BibitemOpen
  \bibinfo {editor} {\bibfnamefont {D.~B.}\ \bibnamefont {Litvin}},\ ed.,\
  \href {https://doi.org/10.1107/9780955360220001} {\emph {\bibinfo {title}
  {Magnetic {{Group Tables}}: 1-, 2- and 3-Dimensional Magnetic Subperiodic
  Groups and Magnetic Space Groups}}}\ (\bibinfo  {publisher} {International
  Union of Crystallography},\ \bibinfo {address} {Chester, England},\ \bibinfo
  {year} {2013})\BibitemShut {NoStop}%
\bibitem [{\citenamefont {{Perez-Mato}}\ \emph {et~al.}(2015)\citenamefont
  {{Perez-Mato}}, \citenamefont {Gallego}, \citenamefont {Tasci}, \citenamefont
  {Elcoro}, \citenamefont {De~La~Flor},\ and\ \citenamefont
  {Aroyo}}]{perez-mato_symmetrybased_2015}%
  \BibitemOpen
  \bibfield  {author} {\bibinfo {author} {\bibfnamefont {J.}~\bibnamefont
  {{Perez-Mato}}}, \bibinfo {author} {\bibfnamefont {S.}~\bibnamefont
  {Gallego}}, \bibinfo {author} {\bibfnamefont {E.}~\bibnamefont {Tasci}},
  \bibinfo {author} {\bibfnamefont {L.}~\bibnamefont {Elcoro}}, \bibinfo
  {author} {\bibfnamefont {G.}~\bibnamefont {De~La~Flor}},\ and\ \bibinfo
  {author} {\bibfnamefont {M.}~\bibnamefont {Aroyo}},\ }\bibfield  {title}
  {\bibinfo {title} {Symmetry-{{Based Computational Tools}} for {{Magnetic
  Crystallography}}},\ }\href
  {https://doi.org/10.1146/annurev-matsci-070214-021008} {\bibfield  {journal}
  {\bibinfo  {journal} {Annu. Rev. Mater. Res.}\ }\textbf {\bibinfo {volume}
  {45}},\ \bibinfo {pages} {217} (\bibinfo {year} {2015})}\BibitemShut
  {NoStop}%
\bibitem [{\citenamefont {Togo}\ \emph {et~al.}(2018)\citenamefont {Togo},
  \citenamefont {Shinohara},\ and\ \citenamefont {Tanaka}}]{togo_spglib_2018}%
  \BibitemOpen
  \bibfield  {author} {\bibinfo {author} {\bibfnamefont {A.}~\bibnamefont
  {Togo}}, \bibinfo {author} {\bibfnamefont {K.}~\bibnamefont {Shinohara}},\
  and\ \bibinfo {author} {\bibfnamefont {I.}~\bibnamefont {Tanaka}},\ }\href
  {https://doi.org/10.48550/ARXIV.1808.01590} {\bibinfo {title} {Spglib: A
  software library for crystal symmetry search}} (\bibinfo {year}
  {2018})\BibitemShut {NoStop}%
\bibitem [{\citenamefont {Tan}\ and\ \citenamefont
  {Rappe}(2019)}]{tan_effect_2019}%
  \BibitemOpen
  \bibfield  {author} {\bibinfo {author} {\bibfnamefont {L.~Z.}\ \bibnamefont
  {Tan}}\ and\ \bibinfo {author} {\bibfnamefont {A.~M.}\ \bibnamefont
  {Rappe}},\ }\bibfield  {title} {\bibinfo {title} {Effect of wavefunction
  delocalization on shift current generation},\ }\href
  {https://doi.org/10.1088/1361-648X/aaf74b} {\bibfield  {journal} {\bibinfo
  {journal} {J. Phys.: Condens. Matter}\ }\textbf {\bibinfo {volume} {31}},\
  \bibinfo {pages} {084002} (\bibinfo {year} {2019})}\BibitemShut {NoStop}%
\bibitem [{\citenamefont {Kaner}\ \emph {et~al.}(2020)\citenamefont {Kaner},
  \citenamefont {Wei}, \citenamefont {Jiang}, \citenamefont {Li}, \citenamefont
  {Xu}, \citenamefont {Pang}, \citenamefont {Li}, \citenamefont {Yang},
  \citenamefont {Jiang}, \citenamefont {Zhang},\ and\ \citenamefont
  {Tian}}]{kaner_enhanced_2020}%
  \BibitemOpen
  \bibfield  {author} {\bibinfo {author} {\bibfnamefont {N.~T.}\ \bibnamefont
  {Kaner}}, \bibinfo {author} {\bibfnamefont {Y.}~\bibnamefont {Wei}}, \bibinfo
  {author} {\bibfnamefont {Y.}~\bibnamefont {Jiang}}, \bibinfo {author}
  {\bibfnamefont {W.}~\bibnamefont {Li}}, \bibinfo {author} {\bibfnamefont
  {X.}~\bibnamefont {Xu}}, \bibinfo {author} {\bibfnamefont {K.}~\bibnamefont
  {Pang}}, \bibinfo {author} {\bibfnamefont {X.}~\bibnamefont {Li}}, \bibinfo
  {author} {\bibfnamefont {J.}~\bibnamefont {Yang}}, \bibinfo {author}
  {\bibfnamefont {Y.}~\bibnamefont {Jiang}}, \bibinfo {author} {\bibfnamefont
  {G.}~\bibnamefont {Zhang}},\ and\ \bibinfo {author} {\bibfnamefont {W.~Q.}\
  \bibnamefont {Tian}},\ }\bibfield  {title} {\bibinfo {title} {Enhanced
  {{Shift Currents}} in {{Monolayer 2D GeS}} and {{SnS}} by {{Strain-Induced
  Band Gap Engineering}}},\ }\href {https://doi.org/10.1021/acsomega.0c01319}
  {\bibfield  {journal} {\bibinfo  {journal} {ACS Omega}\ }\textbf {\bibinfo
  {volume} {5}},\ \bibinfo {pages} {17207} (\bibinfo {year}
  {2020})}\BibitemShut {NoStop}%
\end{thebibliography}%
\end{document}